\documentclass[final,3p,times]{elsarticle}

\usepackage{graphicx}

\usepackage{amssymb}
\usepackage{amsmath}
\usepackage{subfig}
\usepackage{graphicx}
\usepackage{url}
\usepackage{algpseudocode}
\usepackage{algorithm}
\usepackage{multirow}
\usepackage{wasysym}
\usepackage{marvosym}
\usepackage{color, xcolor, colortbl}
\usepackage{array, booktabs}
\usepackage{xspace}
\usepackage{enumitem}

\usepackage{bm}

\newcommand{\secref}[1]{Section \ref{#1}}
\newcommand{\fref}[1]{Fig.~\ref{#1}}
\newcommand{\tref}[1]{Table~\ref{#1}}

\renewcommand{\eqref}[1]{Eq.~(\ref{#1})}
\newcommand{\eref}[1]{(\ref{#1})}

\newcommand{\mat}[1]{\textrm{\textbf{#1}}}

\newcommand{\psif}{\bm{\psi} (\bm{r}, \bm{\Omega})}
\newcommand{\psifd}{\bm{\psi} (\bm{r}, \bm{\Omega}')}

\newcommand{\xten}[1]{$\times$ 10$^{\text{#1}}$}
\newcommand{\xtenm}[1]{\times \text{10}^{\text{#1}}}

\definecolor{light}{rgb}{0.8,0.8,0.8}
\definecolor{medium}{rgb}{0.6,0.6,0.6}
\definecolor{dark}{rgb}{0.4,0.4,0.4}
\definecolor{darkmed}{rgb}{0.3,0.3,0.3}
\definecolor{darkest}{rgb}{0.2,0.2,0.2}
\definecolor{Black}{rgb}{0,0,0}
\definecolor{White}{rgb}{1,1,1}
\definecolor{lightpurple}{rgb}{0.78823,0.709803,0.74509}
\definecolor{lightpurpletext}{rgb}{0.788235,0.5529411,0.658823}
\definecolor{skyblue}{rgb}{0.80392,0.866666,0.92941}
\definecolor{skybluetext}{rgb}{0.61568627,0.7647058,0.913725}
\definecolor{darkgreen}{rgb}{0.3137254,0.458823,0.18431}
\definecolor{foliagegreen}{rgb}{0.188,0.415,0.105}
\definecolor{steelbluegrey}{rgb}{0.1961,0.2353,0.2392}
\definecolor{highlightblue}{rgb}{0.4078,0.6431,0.85}
\definecolor{matlabblue}{rgb}{0,0.2705,0.85}
\definecolor{darkred}{rgb}{0.8,0.1725,0}
\definecolor{fireenginered}{rgb}{0.505,0.1411,0}
\definecolor{darkpurple}{rgb}{0.6431,0.3137,0.8509}
\definecolor{gaylordpurple}{rgb}{0.416,0.204,0.549}
\definecolor{deludedorange}{rgb}{0.7409,0.4392,0}
\definecolor{darksalmon}{rgb}{0.9137,0.411,0.706}

\newcolumntype{a}{>{\columncolor{light}}c}
\newcolumntype{b}{>{\columncolor{skyblue}}c}

\begin{document}

\begin{frontmatter}



\title{AIR multigrid with GMRES polynomials (AIRG) and additive preconditioners for Boltzmann transport\tnoteref{crown}}
\author[AMCG]{S. Dargaville}
\ead{dargaville.steven@gmail.com}
\tnotetext[crown]{UK Ministry of Defence © Crown owned copyright 2023/AWE}
\author[AWE,AMCG]{R.P. Smedley-Stevenson}
\author[AMEC,AMCG]{P.N. Smith}
\author[AMCG]{C.C. Pain}
\address[AMCG]{Applied Modelling and Computation Group, Imperial College London, SW7 2AZ, UK}
\address[AWE]{AWE, Aldermaston, Reading, RG7 4PR, UK}
\address[AMEC]{ANSWERS Software Service, Jacobs, Kimmeridge House, Dorset Green Technology Park, Dorchester, DT2 8ZB, UK}
\begin{abstract}
We develop a reduction multigrid based on approximate ideal restriction (AIR) for use with asymmetric linear systems. We use fixed-order GMRES polynomials to approximate $\mat{A}_\textrm{ff}^{-1}$ and we use these polynomials to build grid transfer operators and perform F-point smoothing. We can also apply a fixed sparsity to these polynomials to prevent fill-in.  

When applied in the streaming limit of the Boltzmann Transport Equation (BTE), with a P$^0$ angular discretisation and a low-memory spatial discretisation on unstructured grids, this ``AIRG'' multigrid used as a preconditioner to an outer GMRES iteration outperforms the lAIR implementation in \textit{hypre}, with two to three times less work. AIRG is very close to scalable; we find either fixed work in the solve with slight growth in the setup, or slight growth in the solve with fixed work in the setup when using fixed sparsity. Using fixed sparsity we see less than 20\% growth in the work of the solve with either 6 levels of spatial refinement or 3 levels of angular refinement. In problems with scattering AIRG performs as well as lAIR, but using the full matrix with scattering is not scalable. 

We then present an iterative method designed for use with scattering which uses the additive combination of two fixed-sparsity preconditioners applied to the angular flux; a single AIRG V-cycle on the streaming/removal operator and a DSA method with a CG FEM. We find with space or angle refinement our iterative method is very close to scalable with fixed memory use.
\end{abstract}

\end{frontmatter}
\section{Introduction}
\label{sec:Introduction}
The Boltzmann transport equation (BTE) describes the distribution of particles moving through an interacting medium and is used to model radiation transport, along with spectral-waves and fluid problems through kinetic and lattice-Boltzmann methods. The mono-energetic steady-state form of the Boltzmann Transport Equation (BTE), with linear scattering and straight-line propagation is written in \eref{eq:bte} as
\begin{equation}
\bm{\Omega} \cdot \nabla_{\bm{r}} \psif + \sigma_\textrm{t} \psif - \int_{\bm{\Omega}'} \sigma_\textrm{s} (\bm{r}, \bm{\Omega}' \rightarrow \bm{\Omega}) \psifd \textrm{d}\bm{\Omega}'  = S_{\textrm{e}}(\bm{r}, \bm{\Omega}).
\label{eq:bte}
\end{equation}
Equation \eref{eq:bte} is a 5-dimensional linear PDE, with three spatial dimensions and two angular dimensions; we neglect the energy and time dimensions. The angular flux, $\psif$, describes the number of particles moving in direction $\bm{\Omega}$, at spatial position $\bm{r}$. The macroscopic total cross section of the material that the particles are moving through is given by $\sigma_\textrm{t}$, which describes particles removed either through absorption or scattering in the material. The source of particles coming from scattering in many radiative processes is given by an integral term, where $\sigma_\textrm{s}$ is the macroscopic scatter cross-sections for this process that describes how particles scatter from direction $\bm{\Omega}'$ into direction $\bm{\Omega}$. Finally any external sources of particles are given by $S_\textrm{e}$. 

One of the main challenges in solving \eref{eq:bte} is that when the scattering cross-sections are large (i.e., the particles are interacting with the material), the BTE tends to a diffusion equation, whereas when the scattering and total cross-sections are zero (i.e., particles are moving through a vacuum), the BTE is purely hyperbolic and hence stable discretisations must be used and the resulting linear systems are asymmetric and non-normal.

In \eref{eq:bte}, the scattering cross-section, $\sigma_\textrm{s}$, for any given angle-to-angle scattering event is often described by a Legendre expansion whose coefficients we denote as $\bm{\sigma}_\textrm{s}$. If we discretise in space/angle (with an angular discretisation like S$_n$), we denote the angular flux as $\bm{\psi}$ and the angular flux in Legendre space as $\bm{\varphi}$. We then write \eref{eq:bte} as a $2\times2$ block system, namely
\begin{equation}
\begin{bmatrix}
\mat{I} & \mat{D}_\textrm{m} \\
-\mat{M}_\textrm{m} \bm{\Sigma}_\textrm{s} & \mat{L}
\end{bmatrix}
\begin{bmatrix}
\bm{\varphi} \\
\bm{\psi}
\end{bmatrix} = 
\begin{bmatrix}
\bm{0} \\
\hat{\mat{S}}_\textrm{e}
\end{bmatrix},
\label{eq:block_bte}
\end{equation}
where $\mat{L}$ is the streaming/removal operator, $\bm{\Sigma}_\textrm{s}$ is a matrix with the scattering cross-sections for each spatial node, $\mat{D}_\textrm{m}$ and $\mat{M}_\textrm{m}$ are formed from the tensor product of the spatial mass matrices and the mapping operators which map between our angular discretisation and the moments of the Legendre space and $\hat{\mat{S}}_\textrm{e}$ is the discretised source term from \eref{eq:bte}. When discretised in space with an upwind discretisation and an appropriate ordering of $\bm{\psi}$ is applied, $\mat{L}$ is a block-diagonal matrix, where each of the blocks is lower triangular and corresponds to the spatial coupling for each direction (i.e., each direction in $\mat{L}$ is not coupled to the others; $\mat{L}$ represents a set of advection equations for each direction).

Equation \eref{eq:block_bte} is typically solved by forming the Schur complement of block $\mat{L}$, namely
\begin{equation}
(\mat{I} + \mat{D}_\textrm{m} \mat{L}^{-1} \mat{M}_\textrm{m} \bm{\Sigma}_\textrm{s}) \bm{\varphi} = -\mat{D}_\textrm{m} \mat{L}^{-1} \hat{\mat{S}}_{\textrm{e}},
\label{eq:schur}
\end{equation}
and then a preconditioned Richardson iteration (known as a source iteration in the transport community), with preconditioner $\mat{M}^{-1}$ is applied to recover
\begin{equation}
\bm{\varphi}^{n+1} = \bm{\varphi}^{n} - \mat{M}^{-1}(\mat{D}_\textrm{m} \mat{L}^{-1}(\mat{M}_\textrm{m} \bm{\Sigma}_\textrm{s} \bm{\varphi}^{n} + \hat{\mat{S}}_{\textrm{e}}) + \bm{\varphi}^{n}).
\label{eq:richard}
\end{equation}
Computing the solution to \eref{eq:block_bte} therefore only requires the Legendre representation of the angular flux (the angular flux can easily be formed, either at a single spatial point or across the domain if needed), as each of the components of \eref{eq:richard} are block-diagonal. This allows for a very low memory iterative method. If there is scattering in the problem, then typically an additive preconditioner like $\mat{M}^{-1}=\mat{I} + \mat{D}_\textrm{diff}^{-1}$ is used; this is known as diffusion-synthetic acceleration (DSA), where $\mat{D}_\textrm{diff}$ is a diffusion operator, with diffusion/sink/boundary coefficients taken from an asymptotic analysis of the BTE in the diffusion limit. 

The spatial discretisation applied to the diffusion operator in DSA can govern its effectiveness in accelerating convergence; research into discretisations of the diffusion operator that are effective and/or ``consistent'' with transport are extensive. The use of a Krylov method to solve \eref{eq:schur} instead of a Richardson method can reduce the dependence on this consistency \cite{Warsa2004} and hence ``inconsistent'' discretisations can be applied to the diffusion operator as part of a DSA resulting in SPD systems that can be solved efficiently. Transport synthetic acceleration (TSA) \cite{Ramone1997} has also been used to accelerate convergence in the scattering limit, where lower order transport solutions are used instead of diffusion; see \cite{Adams2002} for a review of both DSA and TSA method in transport. 

The solution of \eref{eq:schur} relies on the exact inversion of $\mat{L}$; if as mentioned above the spatial and angular discretisations result in $\mat{L}$ with lower-triangular structure then $\mat{L}$ can be inverted exactly with a single (matrix-free) Gauss-Seidel (GS) iteration, also known as a sweep in the transport community. If unstructured spatial grids are used, inverting $\mat{L}$ can become more difficult and finding an ideal sweep ordering is then NP-complete. Furthermore, the introduction of different spatial or angular discretisations, parallelism, time dependence or additional physics that cannot be mapped to Legendre space compounds this problem. Similarly, if we wish to use angular adaptivity where the angular resolution differs throughout the spatial grid, $\mat{L}$ is no longer block-diagonal.

Our goal in the AMCG has been to investigate different discretisations, adaptive and iterative methods for solving the BTE, that have the potential to overcome these difficulties while remaining scalable (i.e., perform a fixed amount of work with fixed memory consumption, even with space/angle refinement). We instead form the Schur complement of block $\mat{I}$ from \eref{eq:block_bte} and solve the system formed with the angular flux, namely
\begin{equation}
(\mat{L} + \mat{M}_\textrm{m} \bm{\Sigma}_\textrm{s} \mat{D}_\textrm{m}) \bm{\psi} = \hat{\mat{S}}_{\textrm{e}}.
\label{eq:schur_us}
\end{equation}
There are several disadvantages to solving \eref{eq:schur_us} instead of \eref{eq:schur}, the most important of which is the considerable increase in memory required. The angular flux, $\bm{\psi}$, is much bigger than $\bm{\varphi}$ and the matrix $\mat{L} + \mat{M}_\textrm{m} \bm{\Sigma}_\textrm{s} \mat{D}_\textrm{m}$ is not block diagonal; instead the scattering operator $\mat{M}_\textrm{m} \bm{\Sigma}_\textrm{s} \mat{D}_\textrm{m}$ couples different angles together and results in dense angle-angle blocks where the nnzs scale like $\mathcal{O}(n^2)$ with angle size. These disadvantages would typically preclude the development of a practical transport algorithm. Previously we have tackled those problems through the combination of: using a stable spatial discretisation based on static condensation that has the stencil of a CG discretisation (reducing the size of $\bm{\psi}$ but at the cost of making the blocks in $\mat{L}$ no longer lower triangular), using angular adaptivity to only focus angular resolution where required (reducing the size of $\bm{\psi}$), and using a matrix-free multigrid to solve \eref{eq:schur_us} that does not rely on the explicit construction of $\mat{M}_\textrm{m} \bm{\Sigma}_\textrm{s} \mat{D}_\textrm{m}$ or on the lower triangular structure in $\mat{L}$. 

We showed previously \cite{Dargaville2019, Dargaville2019b} that these techniques perform well on many transport problems, allowing the practical use of both angular adaptivity with high levels of refinement and unstructured spatial grids. These methods do not use GS/sweep smoothers however and do not scale well in the streaming limit where $\bm{\sigma}_\textrm{s}$ tends to zero. Similarly, most multigrid methods in the literature that achieve good performance for the BTE have relied on either block-based smoothers, often on a cell/element, which do not scale with increasing angle size but which perform well in the scattering limit \cite{Manteuffel1994, Lansrud2005, Kanschat2014, Densmore2016}, or GS/sweeps as smoothers which perform well in the streaming limit \cite{Nowak1988, Morel1991, Manteuffel1995, Manteuffel1996, Pautz1999, Chang2007, Lee2010a, Lee2010, Lee2012, Gao2012, Turcksin2012, buchan_sub-grid_2012, Slaybaugh2013, Slaybaug2015, Drumm2017, Lathouwers2019} but as mentioned can be difficult to parallelise on unstructured grids. Developing multigrid methods which perform well with local smoothers is difficult, particularly as hyperbolic problems have long proved difficult to solve with multigrid due to the lack of developed theory for non-SPD matrices. Recently, reduction multigrid methods based on approximate ideal restrictors (AIR) have been developed \cite{Manteuffel2019, Manteuffel2019a, Southworth2017, Hanophy2020} that show good convergence in asymmetric problems, and in particular with the BTE when using point-based smoothers \cite{Hanophy2020}. 

The aim of this work is hence to build an iterative method that solves \eref{eq:schur_us} with the following features:
\begin{enumerate}
\item Compatible with the space/angle discretisations we have developed previously \cite{Dargaville2019, Dargaville2019b} and hence does not rely on $\mat{L}$ having block diagonal and/or lower triangular structure
\item Does not require the explicit construction of $\mat{M}_\textrm{m} \bm{\Sigma}_\textrm{s} \mat{D}_\textrm{m}$
\item Does not require the use of GS/sweeps
\item Scalable in both the streaming and scattering limits with space/angle refinement
\item Compatible with angular adaptivity
\item Good strong and weak scaling in parallel with unstructured grids  
\end{enumerate}
We examine the first four of these points in this paper, and leave investigation of adaptivity and the parallel performance of our method to future work. Here we present two contributions: the first is an algaebraic multigrid method based on combining AIR with low-order GMRES polynomials, which we call AIRG; the second is an iterative method where we apply additive preconditioners to an outer GMRES iteration on the angular flux, based on the streaming/removal operator and a diffusion operator (DSA). Both these contributions can be used independently of the other; for example we could use AIRG to invert $\mat{L}$ and/or the DSA operator as part of a typical DG FEM/source iteration.

We then compare our iterative method to two different reduction multigrids, namely the \textit{hypre} implementation of lAIR and one where sparse-approximate inverses (SAIs) are used instead of our GMRES polynomials and find performance advantages. We therefore have an iterative method on unstructured grids which never requires the assembly of the full matrix in \eref{eq:schur_us}, that has both good performance and fixed memory consumption across all parameter regimes with space/angle refinement for Boltzmann transport problems. 
\section{Discretisations}
\label{sec:Discretisations}
We begin by presenting the spatial and angular discretisations used in this work; they are based on those presented in \cite{hughes_variational_1998, hughes_multiscale_2006, candy_subgrid_2008, buchan_inner-element_2010, Dargaville2019, Dargaville2019b} and hence we only discuss their key features.
\subsection{Angular discretisation}
\label{sec:ang_discs}
We use a P$^0$ DG FEM in angle (or equivalently a cell-centred FVM) with constant area azi/polar elements that we normalise so that the angular mass matrix is the identity. The first level of our angular discretisation is denoted level 1, with one constant basis function per octant. Each subsequent level of refinement comes from splitting an angular element into four at the midpoint of the azi/cosine polar bounds; this is structured, nested refinement. This is equivalent to the Haar wavelet discretisations discussed in \cite{Dargaville2019}; indeed as part of the matrix-free iterative method in \cite{Dargaville2019} an $\mathcal{O}(n)$ mapping to/from this P$^0$ space to Haar space was performed during every matvec. We instead choose to solve in P$^0$ space as we can form an assembled copy of the streaming/removal matrix that has fixed sparsity with angular refinement; this property was not needed for the matrix-free methods used in our previous work, all we required was a scalable mapping between P$^0$ and Haar spaces. We can also adapt in this P$^0$ space in the same manner as our wavelet space \cite{Dargaville2019}, given the equivalence. We investigate adapting in P$^0$ space in future work.
\subsection{Spatial discretisation}
\label{sec:sub-grid}
Our spatial discretisation is a sub-grid scale FEM, which represents the angular flux as $\bm{\psi} = \bm{\phi} + \bm{\theta}$, where $\bm{\phi}$ is the solution on a ``coarse'' scale and $\bm{\theta}$ is the solution on a ``fine'' scale. The finite element expansions for both the fine and coarse scales can be written as
\begin{equation}
\phi(\bm{r}, \bm{\Omega}) \approx \sum_{i=1}^{\eta_N} N_i(\bm{r}) \tilde{\phi}_i(\bm{\Omega}); \qquad \theta(\bm{r}, \bm{\Omega}) \approx \sum_{i=1}^{\eta_Q} Q_i(\bm{r}) \tilde{\theta}_i(\bm{\Omega}),
\label{eq:space}
\end{equation}
with $\eta_N$ continuous basis functions, $N_i$, and $\eta_Q$ discontinuous basis functions, $Q_i$, with $\tilde{\phi}_i$ and $\tilde{\theta}_i$ the expansion coefficients, respectively. In this work we use linear basis functions for both the continuous and discontinuous spatial expansions. 

As described in \secref{sec:ang_discs}, we use a P$^0$ discretisation, with (constant) basis functions $G_j(\bm{\Omega})$, with a spatially varying number of angular elements $\eta_A^i$ and $\eta_D^i$ on the coarse and fine scales respectively (we enforce that DG nodes have the same angular expansion as their CG counterparts). The expansion coefficients $\tilde{\phi}_i$ and $\tilde{\theta}_i$ in \eref{eq:space} can then be written as space/angle expansion coefficients $\tilde{\phi}_{i,j}$ and $\tilde{\theta}_{i,j}$ via the expansion 
\begin{equation}
\tilde{\phi}_i(\bm{\Omega}) \approx \sum_{j=1}^{\eta_A^i} G_j(\bm{\Omega}) \tilde{\phi}_{i,j}; \qquad \tilde{\theta}_i(\bm{\Omega}) \approx \sum_{j=1}^{\eta_D^i} G_j(\bm{\Omega}) \tilde{\theta}_{i,j}.
\label{eq:angle}
\end{equation}
Following standard FEM theory the discretised form of \eref{eq:bte} can then be written as
\begin{equation}
\begin{bmatrix}
\mat{A} & \mat{B} \\
\mat{C} & \mat{D} \\
\end{bmatrix}
\begin{bmatrix}
\bm{\Phi} \\
\bm{\Theta} \\
\end{bmatrix}
=
\begin{bmatrix}
\mat{S}_{\bm{\Phi}} \\
\mat{S}_{\bm{\Theta}} \\
\end{bmatrix},
\label{eq:SGS_full}
\end{equation}
where ${\bm{\Phi}}$ and ${\bm{\Theta}}$ are vectors containing the coarse and fine scale expansion coefficients, we denote the number of unknowns in ${\bm{\Phi}}$ as NCDOFs and in ${\bm{\Theta}}$ as NDDOFs. The discretised source terms for both scales are $\mat{S}_{\bm{\Phi}}$ and $\mat{S}_{\bm{\Theta}}$. We should note the matrices $\mat{A}$ and $\mat{D}$ are the standard CG and DG FEM matrices that result from discretising \eref{eq:bte}. 

We can then form a Schur complement of block $\mat{D}$ and recover
\begin{equation}
(\mat{A} - \mat{B} \mat{D}^{-1} \mat{C}) {\bm{\Phi}} = \mat{S}_{\bm{\Phi}} - \mat{B} \mat{D}^{-1} \mat{S}_{\bm{\Theta}}.
\label{eq:SGS}
\end{equation}
The fine solution $\bm{\Theta}$ can then be computed
\begin{equation}
\bm{\Theta} = \mat{D}^{-1} (\mat{S}_{\bm{\Theta}} - \mat{C} \bm{\Phi}),
\label{eq:theta}
\end{equation}
and our discrete solution is the addition of both the coarse and fine solutions, namely $\bm{\Psi} = \bm{\Phi} + \bm{\Theta}$ (where the coarse solution $\bm{\Phi}$ has been projected onto the fine space). In order to solve \eref{eq:SGS} and \eref{eq:theta} efficiently/scalably, we must sparsify $\mat{D}$ and this sparsification is dependent on the angular discretisation used, see \cite{buchan_inner-element_2010, Goffin2014, Dargaville_2014, Goffin2015a, Buchan2016, Adigun2018, Dargaville2019, Dargaville2019a} for examples. In this work we replace $\mat{D}^{-1}$ in \eref{eq:SGS} and \eref{eq:theta} with $\hat{\mat{D}}^{-1}$, which is the streaming operator with removal and self-scatter only, and vacuum conditions applied on each DG element (as this removes the jump terms that couple the DG elements, resulting in element blocks). We can then invert this matrix element-wise and store the result, with a constant nnz fraction with space/angle refinement, as it has the same stencil as the streaming operator. Now if we consider the streaming/removal (denoted with a subscript $\Omega$) and scattering contributions (denoted with a subscript S) in \eref{eq:SGS} separately, along with our sparsified $\hat{\mat{D}}^{-1}$ we can rearrange \eref{eq:SGS} and write
\begin{equation}
\left(\mat{A}_\Omega - \mat{B}_\Omega \hat{\mat{D}}^{-1} \mat{C}_\Omega \right) {\bm{\Phi}} + \left((\mat{A}_\textrm{S} + \mat{B}_\textrm{S}(y + \hat{\mat{D}}^{-1} \mat{C}_\Omega) + \mat{B}_\Omega y \right) {\bm{\Phi}} = \mat{S}_{\bm{\Phi}} - (\mat{B}_\Omega+ \mat{B}_\textrm{S}) \hat{\mat{D}}^{-1} \mat{S}_{\bm{\Theta}}.
\label{eq:SGS_sep_stream}
\end{equation}
where $y = \hat{\mat{D}}^{-1} \mat{C}_\textrm{S}$ and our fine component is $\bm{\Theta} = \hat{\mat{D}}^{-1} (\mat{S}_{\bm{\Theta}} - (\mat{C}_\Omega + \mat{C}_\textrm{S}) \bm{\Phi})$. We can see that \eref{eq:SGS_sep_stream} is now written similarly to \eref{eq:schur_us}, where the left term is (very close to) the sub-grid scale streaming/removal operator, although as mentioned in \secref{sec:Introduction} it does not feature lower triangular blocks. We should note that the left term is not exactly the streaming/removal operator as $\hat{\mat{D}}^{-1}$ contains self-scatter, but we call it such below for simplicity; practically we can modify our stabilisation such that $\hat{\mat{D}}^{-1}$ does not contain self-scatter without any substantial differences to either the stability of our system or the preconditioning described in \secref{sec:iterative}. We cannot explicitly form the scattering contribution as it has dense blocks, but given that $\hat{\mat{D}}^{-1}$ has fixed sparsity and the scatter contributions from $\mat{A}_\textrm{S}, \mat{B}_\textrm{S}$ and $\mat{C}_\textrm{S}$ can be formed in Legendre space and mapped back, we can scalably compute a matrix-free matvec with angular refinement in P$^0$ space (with a fixed scatter order). 

There are many ways to implement a matvec for \eref{eq:SGS_sep_stream} (these involve further rearrangement of \eref{eq:SGS_sep_stream}), depending on the amount of memory available. Some key points include that away from boundaries, the element matrices in $\mat{A}_\Omega$, $\mat{B}_\Omega$, $\mat{C}_\Omega$ and $\hat{\mat{D}}^{-1}$ all share the same (block) sparsity; in fact given matching angular resolution and the same order of basis functions on the continuous and discontinuous spatial meshes (we enforce both conditions) two of the element matrices are identical, with $\mat{B}_\Omega$ and $\mat{C}_\Omega$ related by the transpose of the spatial table. A matvec with these components could therefore consist of performing several matvecs on small (max $3\times 3$ or $4\times 4$) angular blocks which can be kept in cache (similar to the work performed during a DG sweep, but without the Gauss-Seidel dependency on ordering). We can also combine a number of the maps to/from Legendre space. We describe some of the choices we make in \secref{sec:Results}, but note that a FLOP count shows our sub-grid scale matvec with scattering can be computed with 1.8x the FLOPs of an equivalent DG matvec.  

With our sub-grid scale discretisation, we are therefore choosing to increase the number of (local) FLOPs in order to significantly decrease our memory consumption. We can build iterative methods that only depend on the the coarse-scale solution, ${\bm{\Phi}}$, which is on a CG stencil, which in 3D has approximately $20\times$ fewer unknowns than a DG method; e.g., a GMRES(20) space built on $\bm{\Phi}$ would take approximately the same space as a single copy of $\bm{\Psi}$ in 3D. One additional benefit is as noted in \cite{Dargaville2019}, is that with either a wavelet or non-wavelet angular discretisation, our sub-grid scale discretisation does not require interpolation between areas of different angular resolution (e.g., across faces), due to the lack of DG jump terms.
\section{Additively preconditioned iterative method}
\label{sec:iterative}
This section details the iterative method we use to solve \eref{eq:SGS_sep_stream}. For simplicity we begin by writing the unseparated form \eref{eq:SGS} with our sparsified $\hat{\mat{D}}^{-1}$ and introduce a preconditioner $\mat{M}^{-1}$ on the right to give
\begin{equation}
(\mat{A} - \mat{B}\hat{\mat{D}}^{-1} \mat{C}) \mat{M}^{-1} \mat{u} 	 = \mat{S}_{\bm{\Phi}} - \mat{B} \mat{D}^{-1} \mat{S}_{\bm{\Theta}}, \quad \mat{u} = \mat{M} \bm{\psi}.
\label{eq:schur_us_precon}
\end{equation}
We solve \eref{eq:schur_us_precon} with GMRES and use a matrix-free matvec as described above to compute the action of $(\mat{A} - \mat{B}\hat{\mat{D}}^{-1} \mat{C})$ (and to compute the source and $\bm{\Theta}$). The preconditioner we apply is based on the additive combination of a two-level method in angle and the streaming/removal operator, which we denote as
\begin{equation}
\mat{M}^{-1} = \mat{M}_\textrm{angle}^{-1} + \mat{M}_\Omega^{-1}.
\label{eq:precon_add}
\end{equation}

The first of these uses a DSA type operator; both DSA and TSA can be thought of as additive angular multigrid preconditioners, with DSA forming a two-level multigrid with the coarse level represented by a diffusion equation, whereas TSA can form multiple coarse grids if desired through lower angular resolution transport discretisations (e.g., see \cite{Manteuffel1994, Pautz1999, Lee2010a, Lee2010, Lee2012, Turcksin2012, Lygidakis2014a, Lygidakis2014, Drumm2017, Lathouwers2019}). We found both were very effective, but in this work we use a DSA method which we write as
\begin{equation}
\mat{M}_\textrm{angle}^{-1} = \mat{R}_\textrm{angle} \mat{D}_\textrm{diff}^{-1} \mat{P}_\textrm{angle}, 
\label{eq:dsa}
\end{equation}
where the angular restrict/prolong are simply the mappings between the constant moment (i.e., the 0th order scatter term) and our P$^0$ angular space. $\mat{D}_\textrm{diff}$ is a standard (DSA) diffusion operator with diffusion coefficient of $\kappa = 1/3 \sigma_s$ and Robin conditions on vacuum boundaries that we discretise with a CG FEM; this makes our DSA ``inconsistent'' but we find this performs well with a Krylov method as the outer iteration, as in \cite{Warsa2004}. For further discussions we appeal to the wealth of literature on different acceleration methods. 

The second component of our additive preconditioner is based on the sub-grid scale streaming/removal operator in \eref{eq:SGS_sep_stream}, namely 
\begin{equation}
\mat{M}_\Omega^{-1} = \left(\mat{A}_\Omega - \mat{B}_\Omega \hat{\mat{D}}^{-1} \mat{C}_\Omega \right)^{-1}.
\label{eq:inverse_streamremoval}
\end{equation}
Due to the block diagonal structure of $\hat{\mat{D}}$ described in \secref{sec:sub-grid}, $\mat{A}_\Omega - \mat{B}_\Omega \hat{\mat{D}}^{-1} \mat{C}_\Omega$ has fixed sparsity with space and (uniform) angle refinement; this is the same as the sparsity of the CG streaming/removal operator, $\mat{A}_\Omega$. This means we can compute/store an assembled version of the sub-grid scale streaming/removal operator scalably and hence use matrix-based approaches (like the multigrid described in \secref{sec:airg}) to apply \eref{eq:inverse_streamremoval} and as part of \eref{eq:SGS_sep_stream} when computing a matvec with scattering if desired. 

The additive combination of these two preconditioners is designed to achieve good performance in both the streaming and scattering limits. For problems with both streaming/scattering regions we could apply each of the additive preconditioners only in regions which require it \cite{Southworth2021}. It is trivial to form the grid-transfer operators $\mat{R}_\textrm{angle}$ and $\mat{P}_\textrm{angle}$ regardless of angular refinement, but for a scalable iterative method we must be able to apply the inverses of both the diffusion matrix with a fixed amount of work given spatial refinement, and the streaming/removal operator with fixed work given space/angle refinement. We can do so inexactly given these are applied as preconditioners. This is in contrast to a source iteration like \eref{eq:richard}, where \cite{Hanophy2020} notes that inexact application of $\mat{L}^{-1}$ changes the discrete solution and they show that increasingly accurate solves are required with grid refinement.

The next section describes a novel reduction multigrid that we can use to apply the inverses of our streaming/removal operator, $\mat{A}_\Omega - \mat{B}_\Omega \hat{\mat{D}}^{-1} \mat{C}_\Omega$ (or the streaming operator in the limit of zero total cross-section), and a CG diffusion operator, $\mat{D}_\textrm{diff}$ if desired. For comparison purposes we also test AIRG on the full matrix, $\mat{A} - \mat{B}\hat{\mat{D}}^{-1} \mat{C}$, in the scattering limit; we cannot form this matrix scalably but it helps demonstrate that AIRG is applicable in both advective and diffuse problems. 
\section{AIRG multigrid}
\label{sec:airg}
We begin this section with a summary of a reduction multigrid \cite{Notay2005, MacLachlan2006, Brannick2010a, Southworth2017, Manteuffel2019, Zaman2022}; we should note that reduction multigrids, block LDU factorisations/preconditioners and multi-level ILU methods all use similar building blocks (we discuss this further below). If we consider a general linear system $\mat{A}\mat{x}=\mat{b}$ we can form a block-system due to a coarse/fine (CF) splitting as 
\begin{equation}
\begin{bmatrix}
\mat{A}_\textrm{ff} & \mat{A}_\textrm{fc} \\
\mat{A}_\textrm{cf} & \mat{A}_\textrm{cc}
\end{bmatrix}
\begin{bmatrix}
\bm{x_\textrm{f}} \\
\bm{x_\textrm{c}}
\end{bmatrix} = 
\begin{bmatrix}
\bm{b_\textrm{f}} \\
\bm{b_\textrm{c}}
\end{bmatrix}.
\label{eq:air_two}
\end{equation}
If we write the prolongator and restrictor as 
\begin{equation}
\mat{P} = 
\begin{bmatrix}
\mat{W} \\
\mat{I}
\end{bmatrix}, \quad 
\mat{R} = 
\begin{bmatrix}
\mat{Z} & \mat{I}
\end{bmatrix},
\end{equation}
then we can form a coarse grid matrix as $\mat{A}_\textrm{coarse}=\mat{R}\mat{A}\mat{P}$ and a multgrid hierarchy can be built by applying the same technique to $\mat{A}_\textrm{coarse}$. To see how the operators $\mat{R}$ and $\mat{P}$ are constructed we consider an exact two-grid method, where down smoothing (``pre'') occurs before restriction and up smoothing occurs after prolongation (``post''). We can write the error at the $i$th step as $\mat{e}^i = \bar{\mat{x}}-\mat{x}^i$, where $\bar{\mat{x}}$ is the exact solution to our linear system. Following \cite{Manteuffel2019}, if we have error on the top grid after the down smooth, we can form our coarse grid residual by computing $\mat{R} \mat{A} \mat{e}^i$ and hence the error after a coarse grid solve is $\mat{A}_\textrm{coarse}^{-1} \mat{R} \mat{A} \mat{e}^i$. The error on the top grid after coarse grid correction, denoted as $\mat{e}^{i+1}$ is hence
\begin{equation}
\mat{e}^{i+1} = (\mat{I} - \mat{P} \mat{A}_\textrm{coarse}^{-1} \mat{R} \mat{A}) \mat{e}^i.
\label{eq:coarse_error}
\end{equation}
We can consider the error, $\mat{e}^i$, to be made up of two components, one of which is in the range of interpolation and a remainder, namely
\begin{equation}
\mat{e}^i = 
\begin{bmatrix}
\mat{e}_\textrm{f} \\
\mat{e}_\textrm{c}
\end{bmatrix} = 
\begin{bmatrix}
\mat{W} \mat{e}_\textrm{c} \\
\mat{e}_\textrm{c}
\end{bmatrix} + 
\begin{bmatrix}
\bm{\delta} \mat{e}_\textrm{f}^i \\
\mat{0}
\end{bmatrix}.
\end{equation}
We can then write \eref{eq:coarse_error} as
\begin{equation}
\mat{e}^{i+1} = (\mat{I} - \mat{P} \mat{A}_\textrm{coarse}^{-1} \mat{R} \mat{A}) \begin{bmatrix}
\bm{\delta} \mat{e}_\textrm{f}^i \\
\mat{0}
\end{bmatrix}.
\label{eq:coarse_error_two}
\end{equation}
If $\bm{\delta} \mat{e}_\textrm{f}^i=\mat{0}$ then F-point error is in the range of interpolation and the coarse-grid correction is exact, as $\mat{e}^{i+1}=\mat{0}$. Methods based on ideal restriction however choose $\mat{Z}$ to ensure that
\begin{equation}
\mat{R} \mat{A} 
\begin{bmatrix}
\bm{\delta} \mat{e}_\textrm{f}^i \\
\mat{0}
\end{bmatrix}=
\mat{0}.
\label{eq:ra_zero}
\end{equation}
This enforces that any error on the top grid that is not in the range of interpolation does not make it down to the coarse grid. A local form of \eref{eq:ra_zero} is used explicitly to form the restrictor in lAIR. Expanding \eref{eq:ra_zero} gives the condition 
\[
(\mat{Z} \mat{A}_\textrm{ff} + \mat{A}_\textrm{cf}) \bm{\delta} \mat{e}_\textrm{f}^i = \mat{0},
\]
and hence $\mat{Z}=-\mat{A}_\textrm{cf} \mat{A}_\textrm{ff}^{-1}$ is known as the ``ideal'' restrictor that satisfies this condition. The error after coarse-grid correction \eref{eq:coarse_error_two} then becomes 
\begin{equation}
\mat{e}^{i+1} = 
\begin{bmatrix}
\mat{e}_\textrm{f}^i - \mat{W} \mat{e}_\textrm{c}^i\\
\mat{0}
\end{bmatrix}.
\label{eq:error_end}
\end{equation}
Post F-point smoothing to convergence then gives an exact two-grid method. This is one of the defining characteristics of a reduction multigrid, that ideal restriction ensures that after coarse-grid correction, the C-point error is zero and hence F-point smoothing is appropriate; a similar statement can be used to construct an ``ideal'' prolongator given by $\mat{W}=-\mat{A}_\textrm{ff}^{-1} \mat{A}_\textrm{fc}$. Both \cite{Manteuffel2019, Southworth2017} build approximations to the ideal restrictor and then build a classical one-point prolongator that interpolates each F-point from it's strongest C-point connection with injection; by definition this preserves the constant. \cite{Manteuffel2019} show that this is sufficient to ensure good convergence in advection problems, and \cite{Manteuffel2019a} prove more general criteria on the required operators.  

With the ideal operators the near-nullspace vectors are naturally preserved and we don't have to explicitly provide them (or determine them through adaptive methods \cite{Brezina2004, Brezina2005}); in fact any error modes (near-nullspace or otherwise) that are not in the range of interpolation stay on the top grid and should be smoothed, with the rest being accurately restricted to the coarse grid by construction; if we consider near-nullspace vectors this can be easily demonstrated by considering that if $\mat{A} [\mat{n}_\textrm{f}, \mat{n}_\textrm{c}]^\textrm{T}\approx \mat{0}$, where $[\mat{n}_\textrm{f}, \mat{n}_\textrm{c}]$ is a (partitioned) near-nullspace vector, then with the ideal operators $\mat{A}_\textrm{coarse} \mat{n}_\textrm{c} \approx \mat{0}$. This is in contrast to traditional multigrid methods, which smooth high-frequency modes preferentially on the top grid, with interpolation specifically designed to transfer low-frequency modes (i.e., near-nullspace vectors) to the coarse grid, where they become high-frequency and hence can be smoothed easily. 

\cite{Manteuffel2019} show that the error propagation matrix, $\bm{\epsilon}$ of a reduction multigrid with F point up smooths (and no down smooths) is given by
\begin{equation}
\bm{\epsilon} = \mat{I} - \mat{M}_\textrm{LDU}^{-1} \mat{A},
\end{equation}
where $\mat{M}_\textrm{LDU}$ is a block LDU factorisation of $\mat{A}$ given by
\begin{equation}
\mat{M}_\textrm{LDU}
=
\begin{bmatrix}
\mat{I} & \mat{0} \\
\mat{A}_\textrm{cf} \mat{A}_\textrm{ff}^{-1} & \mat{I}
\end{bmatrix}
\begin{bmatrix}
\mat{A}_\textrm{ff} & \mat{0} \\
\mat{0} & \mat{S}
\end{bmatrix}
\begin{bmatrix}
\mat{I} & \mat{A}_\textrm{ff}^{-1} \mat{A}_\textrm{fc} \\
\mat{0} & \mat{I}
\end{bmatrix},
\label{eq:block_ldu}
\end{equation}
where $\mat{S}=\mat{A}_\textrm{cc} - \mat{A}_\textrm{cf} \mat{A}_\textrm{ff}^{-1}\mat{A}_\textrm{fc}$ is the Schur complement of block $\mat{A}_\textrm{ff}$. The inverse of $\mat{M}_\textrm{LDU}$ is given by
\begin{equation}
\mat{M}_\textrm{LDU}^{-1}
=
\begin{bmatrix}
\mat{I} & \mat{W} \\
\mat{0} & \mat{I}
\end{bmatrix}
\begin{bmatrix}
\mat{A}_\textrm{ff}^{-1} & \mat{0} \\
\mat{0} & \mat{S}^{-1}
\end{bmatrix}
\begin{bmatrix}
\mat{I} & \mat{0} \\
\mat{Z} & \mat{I}
\end{bmatrix}.
\label{eq:block_ldu_inverse}
\end{equation}
The ideal operators give that $\mat{A}_\textrm{coarse}=\mat{S}$. Block LDU factorisations formed with approximate ideal operators have a long history of being used as preconditioners. The same factorisation can be applied to $\mat{S}$ to recover a multilevel LDU method. The benefit to using a block LDU method, instead of a reduction multigrid is that up F-point smoothing and coarse grid correction occur additively (as written in \eref{eq:block_ldu_inverse}), rather than F-point smoothing occuring after coarse grid correction with a reduction multigrid (as written in \eref{eq:error_end}). This is an attractive property in parallel. 

If we wish to form a reduction multigrid (or an LDU method) we must approximate $\mat{A}_\textrm{ff}^{-1}$; we denote the approximation as $\hat{\mat{A}}_\textrm{ff}^{-1} \approx \mat{A}_\textrm{ff}^{-1}$ and hence our approximate ideal restrictor and prolongator are given by 
\begin{equation}
\mat{P} = 
\begin{bmatrix}
-\hat{\mat{A}}_\textrm{ff}^{-1} \mat{A}_\textrm{fc} \\
\mat{I}
\end{bmatrix}, \quad
\mat{R} = 
\begin{bmatrix}
-\mat{A}_\textrm{cf} \hat{\mat{A}}_\textrm{ff}^{-1} & \mat{I}
\end{bmatrix}.
\label{eq:prolong}
\end{equation}

The strength of our approximate ideal operators allows the use of zero ``down'' smoothing iterations, with F-point smoothing only on the ``up'' cycle after coarse-grid correction. The authors in \cite{Manteuffel2019} however do perform some C-point smoothing on their ``up'' cycle (with F-F-C Jacobi) for robustness, but we did not find that necessary in this work.

Equation \eref{eq:error_end} suggests a simple choice for the prolongator would be $\mat{W}=\mat{0}$, but in a multi-level setting where we are approximating $\mat{A}_\textrm{ff}^{-1}$ \cite{Southworth2017, Manteuffel2019} show this would require an increasingly accurate approximation with grid refinement. Rather than build a classical prolongator like \cite{Manteuffel2019, Southworth2017}, as we have an approximation of $\mat{A}_\textrm{ff}^{-1}$ it only costs one extra matmatmult per level to compute the ideal prolongator, $\mat{P}$. We then keep only the largest (in magnitude) entry, and hence we form a one-point ideal prolongator. This is like using AIR in conjunction with AIP which \cite{Manteuffel2019a} suggest could form a scalable method for non-symmetric systems. This is different to lAIR \cite{Southworth2017}, where $\mat{Z}$ is computed directly; computing the ideal prolongator would therefore require an equivalent calculation for $\mat{W}$. We find this is a robust choice for our prolongator (as it satisfies the approximation properties required by \cite{Manteuffel2019a}) while also being very simple as it doesn't require knowledge of the near-nullspace.

To perform up F-point smoothing on each level, we use a Richardson iteration to apply $\hat{\mat{A}}_\textrm{ff}^{-1}$. On each level if we are smoothing $\mat{A}\mat{e}=\mat{r}$, where $\mat{r}$ is the residual computed after the coarse-grid correction then
\begin{equation}
\mat{e}_\textrm{f}^{n+1} = \mat{e}_\textrm{f}^{n} + \hat{\mat{A}}_\textrm{ff}^{-1}(\mat{r}_\textrm{f} - \mat{A}_\textrm{fc} \mat{e}_\textrm{c}^{n} - \mat{A}_\textrm{ff} \mat{e}_\textrm{f}^{n}). 
\label{eq:f_point}
\end{equation}
The coarse-grid error does not change during this process, so we can cache the result of $\mat{A}_\textrm{fc} \mat{e}_\textrm{c}^{n}$ during multiple F-point smooths. The next section details how we construct our approximation $\hat{\mat{A}}_\textrm{ff}^{-1}$.
\subsection{GMRES polynomials}
\label{sec:frozen}
Forming good, but sparse approximations to $\mat{A}_\textrm{ff}^{-1}$ is the key to a reduction multigrid (and LDU methods as mentioned); this is achieved in-part by ensuring a ``good'' CF splitting that results in a well-conditioned $\mat{A}_\textrm{ff}$. In particular $\mat{A}_\textrm{ff}$ is often better conditioned (or more diagonally dominant) than $\mat{A}$. In this section we assume a suitable CF splitting has been performed; see \secref{sec:F and C point selection} for more details.  

The original AMGr work \cite{MacLachlan2006} approximated $\mat{A}_\textrm{ff}^{-1}$ using the inverse diagonal of $\mat{A}_\textrm{ff}$. The nAIR method presented by \cite{Manteuffel2019} on advection-diffusion equations used a matrix-polynomial approximation of $\mat{A}_\textrm{ff}^{-1}$ generated from a truncated Neumann series; unfortunately this does not converge well in the diffuse limit (i.e., when $\mat{A}_\textrm{ff}$ is not lower-triangular). The work in \cite{Southworth2017} however solved dense, local linear systems, which enforce that $\mat{R}\mat{A}=\mat{0}$ within a certain F-point sparsity pattern to compute each row of $\mat{Z}$ directly; this was denoted as lAIR. For advection-diffusion equations, this showed good performance in both the advection and diffuse limit. The work of \cite{Hanophy2020} specifically examined the performance of lAIR in parallel for the BTE, but used on $\mat{L}$ in \eref{eq:schur}. Both \cite{Chow2003} and \cite{Zaman2022} showed the use of sparse approximate inverses (SAIs) \cite{Kolotilina1993, Grote1997} to approximate $\mat{A}_\textrm{ff}^{-1}$. SAIs solve the minimisation problem, $\textrm{argmin}_{\mat{M}} ||\mat{I} - \mat{A}_\textrm{ff} \mat{M}||_\textrm{F}^2$ (which can be written as a separate least-squares problem for each column, or the left equivalent for each row), to generate an approximate inverse $\mat{M}$; the fixed sparsity of $\mat{A}_\textrm{ff}$ can also be enforced on $\mat{M}$. Both authors also investigated using SAIs to approximate $\mat{W}$ directly by computing $\textrm{argmin}_{\mat{W}} ||\mat{A}_\textrm{ff} \mat{W} + \mat{A}_\textrm{fc}||_\textrm{F}^2$ (with different enforced sparsity patterns); an equivalent calculation for $\mat{Z}$ is hence similar to lAIR enforcing $\mat{R}\mat{A}$ be exactly zero on each row over a given F-point sparsity. For LDU methods, \cite{Vassilevski2008} outlined many of the techniques used; these include using diagonal approximations, or incomplete LU (often referred to as multi-level ILU \cite{Saad2005}) and Cholesky factorisations.

Rather than explicitly approximate $\mat{A}_\textrm{ff}^{-1}$, many multigrid methods such as smoothed aggregation \cite{vanvek_algebraic_1996, Schroder2010, Wiesner2015} or root-node \cite{Manteuffel2017} methods build their prolongation operators by minimising the energy (in an appropriate norm) of their prolongator. This is equivalent to solving $\mat{A} \mat{P}=\mat{0}$ (either column-wise or in a global sense) and given the equivalent construction of \eref{eq:ra_zero} for the prolongator, these methods can therefore converge to the ideal prolongator. One of the benefits of forming the ideal operators through a minimisation process is that constraints on the sparsity of the resulting operator can be applied at all points through this process. In particular, if this sparsity affects the ability to preserve near-nullspace vectors, their preservation can be enforced at each step as the minimisation converges (e.g., see both \cite{Xu2017} for an overview and \cite{Olson2011} for a general application to asymmetric systems).

In this work we compute $\hat{\mat{A}}_\textrm{ff}^{-1}$ with GMRES polynomials \cite{saad_gmres:_1986, Nachtigal1992, Liu2015, Loe2021, Meurant2020}. Polynomial methods have been used in multigrid for many years. Examples include, as mentioned, in nAIR \cite{Manteuffel2019} to approximate $\mat{A}_\textrm{ff}^{-1}$, in AMLI to approximate the inverse of the coarse-grid matrix \cite{Axelsson1989, Axelsson1990} and commonly with Chebychev polynomials used as smoothers, among others. The aim of using GMRES polynomials in this work is to ensure good approximations to $\mat{A}_\textrm{ff}^{-1}$ that don't depend on $\mat{A}_\textrm{ff}$ being SPD, on particular orderings of the unknowns, or diagonal, lower-triangular or block structure, while still allowing good performance in parallel. As such AIRG should be applicable to many common discretisations of advection-diffusion problems, in both limits, or any operators where a suitable splitting can produce a well-conditioned $\mat{A}_\textrm{ff}$ that a GMRES polynomial can approximate. A key difference between this work and previous work with AIR is that we also reuse these polynomials as our F-point smoothers. This allows us to build a very simple and strong multigrid method for asymmetric problems.

We begin by summarising the GMRES method. For a general linear system $\mat{A}\mat{x}=\mat{b}$, where $\mat{A}$ is $n \times n$, we specify an initial guess $\mat{x}^0=\mat{0}$ and hence an initial residual $\mat{r}^0=\mat{b}$. The Krylov subspace of dimension $m$: $\textrm{span}\{\mat{b}, \mat{A}\mat{b}, \mat{A}^2 \mat{b}, \ldots, \mat{A}^{m-1} \mat{b}\}$ is used in GMRES to build an approximate solution. This solution at step $m$ can be written as $\mat{x}^m = q_{m-1}(\mat{A}) \mat{b}$, where $q_{m-1}(\mat{A})$ is a matrix polynomial of degree $m-1$ known as the GMRES polynomial. This is the polynomial that minimises the residual $||\mat{r}^m|| = ||p(\mat{A})\mat{r}^0||_2$ subject to $p(0)=1$, where $p(\mat{A}) = 1-\mat{A} q_{m-1}(\mat{A})$ is known as the residual polynomial. 

The coefficients of $q_{m-1}$ correspond to the required linear combinations of the Krylov vectors and hence a typical GMRES algorithm can be modified to generate them. \cite{Liu2015, Loe2021} discuss the generation and application of these polynomials in detail and care must be taken if high order polynomials are desired. Thankfully we are only concerned with low-order polynomials for use within our multigrid hierarchy and as such consider two different bases.  

As part of a typical GMRES algorithm, we consider a set of orthonormal vectors which form a basis for our subspace, stored in the columns of the matrix $\mat{V}_m$ (i.e., the Arnoldi basis). Similarly the Krylov vectors $\mat{b}, \mat{A}\mat{b}, \ldots$ make up the columns of the matrix $\mat{K}_m$ (i.e., the power basis). Given the vectors in $\mat{V}_m$ and $\mat{K}_m$ span the same space we can write them as linear combinations of each other and hence $\mat{V}_m = \mat{K}_m \mat{C}_m$, where $\mat{C}_m$ is of size $m \times m$. Typically GMRES doesn't store $\mat{C}_m$ but a small modification can be made to store these values at each GMRES step. The approximate solution produced by GMRES is given by $\mat{x}^m = \mat{x}^0 + \mat{V}_m \mat{y}_m$, where $\mat{y}_m$ comes from the solution of the least-squares problem. The coefficients for the polynomial $q_{m-1}$ can then be computed through $(\alpha_0, \ldots, \alpha_{m-1})^\textrm{T}=\mat{C}_m\mat{y}_m$ (as in \cite{Nachtigal1992}).

Equivalently (in exact arithmetic) a QR factorisation of $\mat{K}_{m+1}=\mat{Q}\mat{R}$ can be computed. If we form the submatrix $\tilde{\mat{R}}$ from $\mat{R}$ but without the first column, and note that $\beta=\mat{R}_{1,1}$ then the polynomial coefficients come from the solution of the least-squares problem $(\alpha_0, \ldots, \alpha_{m-1})^\textrm{T}= \textrm{argmin}_{y_m} ||\beta \mat{e}_1 - \tilde{\mat{R}} \mat{y}_m||_2$, where $\mat{e}_1$ is the first column of the $m+1$ identity. GMRES methods normally don't use the power basis directly as it is poorly conditioned when $m\rightarrow \infty$, although this is typically not a concern at low-order; for example \cite{Liu2015} found using the power basis was stable up to 10th order. In either case our GMRES polynomial of degree $m-1$ is given by
\begin{equation}
q_{m-1}(\mat{A}) = \alpha_0 + \alpha_1 \mat{A} + \alpha_2 \mat{A}^2 + \ldots + \alpha_{m-1}\mat{A}^{m-1} \approx \mat{A}^{-1}.
\label{eq:gmres_poly}
\end{equation}
At each step of a GMRES method, the subspace size, $m$, grows and hence the GMRES polynomial changes. Polynomial preconditioning methods typically freeze the size of the subspace and perform a precompute step to generate a GMRES polynomial of a fixed order. This polynomial is then used as a stationary preconditioner, often for GMRES itself. In this work we want to approximate $\mat{A}_\textrm{ff}^{-1}$ on each multigrid level; we use GMRES polynomials with a fixed polynomial order. 

To generate the coefficients for our GMRES polynomials, during our multigrid setup, on each level of our multigrid hierarchy we set $m$ to a fixed value, assign an initial guess of zero and use a random rhs (see \cite{Greenbaum1994, Loe2019, Loe2021}). We can then chose to use either the Arnoldi or power basis with $\mat{A}_\textrm{ff}$ to form our coefficients. The Arnoldi basis requires $m$ steps of the modified GMRES described above; this costs $m$ matvecs along with a number of dot products and norms (i.e., reductions) on each level. 

Instead we could use the power basis, $\mat{K}_{m+1}$, which still requires $m$ matvecs (the setup of a matrix-power kernel may not be worth it given we only need to compute our low-order polynomial coefficients once), but in parallel a tall-skinny QR (TSQR) factorisation can be used to decrease the number of reductions. This relies on QR factorisations of small local blocks  along with a single all-reduce to generate $\mat{R}$; we don't require $\mat{Q}$ to compute our polynomial coefficients and hence the small local $\mat{Q}$ blocks can be discarded. This is equivalent to modifying a communication-avoiding GMRES with $s=1$ \cite{Hoemmen2010, Loe2019} to generate the coefficients. We tested using both the Arnoldi and power basis in this work and they generate the same polynomial coefficients to near round-off error which shows that stability is not a concern at such low orders. Hence we can use the power basis in parallel and the generation of the polynomial coefficients can be considered as communication-avoiding. 

Rather than use a random vector, we could form a GMRES polynomial by using a block method and solving $\mat{Z} \mat{A}_\textrm{ff} = -\mat{A}_\textrm{cf}$. That polynomial would be tailored to computing $\mat{Z}$ which is not desirable given we also use our polynomial to perform F-point smoothing and hence want to smooth all the modes of $\mat{A}_\textrm{ff}$, as discussed above. 

If we wish to use low-order GMRES polynomials to approximate $\mat{A}_\textrm{ff}^{-1}$, we risk not having converged our approximate ideal operators sufficiently. Similarly, the introduction of fixed sparsity, drop tolerances or equivalent in $\hat{\mat{A}}_\textrm{ff}^{-1}$, $\mat{P}$, $\mat{R}$ or $\mat{A}_\textrm{coarse}$ may exacerbate this. We examine the impact of both varying $m$ and sparsifying our operators below in \secref{sec:nullspace}. 
\subsubsection{Fixed sparsity polynomials}
\label{sec:fixed_sparsity}
We would like to explicitly assemble a matrix representation of our polynomials on each level to build the grid-transfer operators in \eref{eq:prolong}. In the limit $m \rightarrow n$, a GMRES polynomial converges to $\mat{A}_\textrm{ff}^{-1}$ exactly. Practically this is impossible to achieve given the storage requirements (and stability/orthogonality issues). Even at low-order however ($m>2$), we would like to avoid the fill-in that comes from the matrix powers. 

To do this, we can construct our polynomials by enforcing a fixed sparsity on each of the matrix powers; for simplicity we chose the sparsity of $\mat{A}_\textrm{ff}$. This is particularly suited to convection operators given that the fill-in should be small. For example, if we consider a third-order polynomial (i.e., $m=4$), and denoting the sparsity pattern of $\mat{A}_\textrm{ff}$ as $S \subset \{(i, j) \, | \, (\mat{A}_\textrm{ff})_{i,j} \neq 0\}$, we enforce that
\begin{equation}
(\tilde{\mat{A}_\textrm{ff}^2})_{i,j}=(\mat{A}_\textrm{ff} \mat{A}_\textrm{ff})_{i,j}, \quad (\tilde{\mat{A}_\textrm{ff}^3})_{i,j}=(\tilde{\mat{A}_\textrm{ff}^2} \mat{A}_\textrm{ff})_{i,j} \quad (i,j) \in S,
\label{eq:sparsity_powers}
\end{equation}
and for $(i,j)$ not in $S$, the entries are zero. This is simply computing $\mat{A}_\textrm{ff}^2$ with no fill-in, using this approximation when computing subsequent matrix-powers and again enforcing no fill-in on the result. Our fixed-sparsity approximation to $\mat{A}_\textrm{ff}^{-1}$ with $m=4$ would then be given by
\begin{equation}
\hat{\mat{A}}_\textrm{ff}^{-1} = \alpha_0 + \alpha_1 \mat{A}_\textrm{ff} + \alpha_2 \tilde{\mat{A}_\textrm{ff}^2} + \alpha_{3} \tilde{\mat{A}_\textrm{ff}^{3}} \approx q_3(\mat{A}_\textrm{ff}).
\label{eq:gmres_poly_approx}
\end{equation}
Fixing the sparsity of the matrix powers reduces the memory consumption of our hierarchy and also allows us to optimise the construction of $\hat{\mat{A}}_\textrm{ff}^{-1}$. For example, with $m=4$ it costs two matrix-matrix additions and two matmatmults, but given the shared sparsity, the additions can be performed quickly and the matmatmults with $m > 2$ can share the same row data required when computing $\mat{A}_\textrm{ff}^2$. In parallel this means we only require the communication of off-processor row data once with $m > 2$. Computing low-order GMRES polynomials with fixed sparsity with the power basis therefore only requires the communication associated with $m$ matvecs, a single all-reduce and if $m>2$ the matmatmult which produces $\mat{A}_\textrm{ff}^2$, regardless of the polynomial order. 

This has the potential to scale well, in contrast to some of the other methods described in \secref{sec:frozen}. We lack the lower triangular structure in our spatial discretisation that makes nAIR effective, although if we did use an upwind DG FEM discretisation in space, both nAIR and our GMRES polynomials would share the same communication pattern in the matmatmults (and a fixed sparsity version of nAIR would also likely perform well in the streaming limit). Using ILU factorisations make parallelisation difficult given the sequential nature of the underlying Gaussian elimination; approximate ILU factorisations have more parallelism \cite{Chow2015, Anzt2018} but they still require triangle solves (which would then also have to be approximated for better performance in parallel, with truncated Neumann series for example). The SAIs used by \cite{Chow2003, Zaman2022} to approximate $\mat{A}_\textrm{ff}^{-1}$ should scale well, given that if the fixed sparsity of $\mat{A}_\textrm{ff}$ is used, then the formation of an approximate inverse requires the off-process rows in the ``shadow'' of the local column sparsity; an ``incomplete'' SAI \cite{Anzt2018a} would only require the same communication as computing $\mat{A}_\textrm{ff}^2$; similar statements can be made for using either SAIs to compute $\mat{Z}$ or $\mat{W}$ directly or by using lAIR (given equivalent sparsity constraints). We must however also consider the local cost of using SAIs/lAIR and their effectiveness (in particular we found we require greater than distance 2 neighbours in lAIR for scalability but the setup time required for this becomes prohibitive); we examine this in \secref{sec:Results}. 
 
With nAIR the approximations $\hat{\mat{A}}_\textrm{ff}^{-1}$ on each level are thrown away once the grid-transfer operators have been built (and with lAIR $\mat{Z}$ is computed directly) and hence different F-point smoothers are used; as mentioned we keep $\hat{\mat{A}}_\textrm{ff}^{-1}$ on each level to use as smoothers instead. The GMRES polynomials are very strong smoothers (which don't require the calculation of any extra dampening parameters) but storing assembled versions of them (even with fixed sparsity) takes extra memory. We examine the total memory consumption of the AIRG hierarchy in \secref{sec:Results}, but note that with fixed sparsity our experiments with pure streaming show that $\hat{\mat{A}}_\textrm{ff}^{-1}$ has approximately $65\%$ as many non-zeros as $\mat{R}$. We could instead throw away the matrix $\hat{\mat{A}}_\textrm{ff}^{-1}$ after our grid transfer operators are computed on each level and perform F-point smoothing by applying $q_{m-1}(\mat{A}_\textrm{ff})$ matrix-free. With $m\le 2$ the number of matvecs required is the same and hence $\hat{\mat{A}}_\textrm{ff}^{-1}$ could be discarded. With $m > 2$ however, applying $q_{m-1}(\mat{A}_\textrm{ff})$ matrix-free would require more matvecs in order to apply the matrix-powers. As such we believe the (constant sized) extra memory required is easily justified. We also investigated using Chebyshev polynomials as smoothers. We precomputed the required eigenvalue estimates with GMRES in order to build the bounding circle/ellipse given our asymmetric linear systems. We found that smoothing with these polynomials was far less efficient/robust (and practically more difficult) than using the GMRES polynomials; indeed using the GMRES polynomials directly is one of the key messages of \cite{Nachtigal1992}. 
\subsubsection{Drop tolerance on $\mat{A}_\textrm{ff}$}
\label{sec:drop_aff}
If our original linear system $\mat{A}\mat{x}=\mat{b}$ is not sparse, neglecting the fill-in with the fixed sparsity polynomials in \secref{sec:fixed_sparsity} may not be sufficient by itself to ensure a practical multigrid method. We can also introduce a drop tolerance to $\mat{A}_\textrm{ff}$ that is applied prior to constructing our polynomial approximations $\hat{\mat{A}}_\textrm{ff}^{-1}$. In general this is not necessary in the streaming limit, but with scattering we find it helps keep the complexity low. This is similar to the strong $\mat{R}$ threshold used in the \textit{hypre} implementation of lAIR, which determines strong neighbours prior to the construction of $\mat{Z}$, and in \secref{sec:Results} we denote it as such. We can either apply the dropping after the polynomial coefficients are computed (in a similar fashion to how the fixed sparsity in \secref{sec:fixed_sparsity} is applied), or before such that we are forming a polynomial approximation to a sparsified $\mat{A}_\textrm{ff}$. With scattering, we did not find much of a difference in convergence so we chose to apply the dropping before, as this helps reduce the cost of the matvec used to compute the polynomial coefficients. 
\subsection{Approximations of $\mat{A}_\textrm{ff}^{-1}$}
\label{sec:nullspace}
\begin{figure}[th]
\centering
\subfloat[][Operators formed from GMRES polynomials without fixed sparsity.]{\label{fig:spectrum_exact}\includegraphics[width =0.45\textwidth]{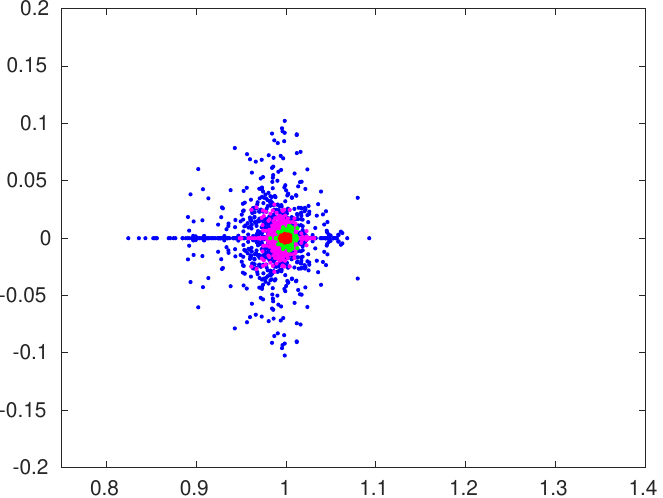}} \hspace{0.1cm}
\subfloat[][Operators formed from fixed sparsity GMRES polynomials and relative drop tolerances applied to the resulting $\mat{R}$ and $\mat{P}$.]{\label{fig:spectrum_coarse_mat_not_sparse}\includegraphics[width =0.45\textwidth]{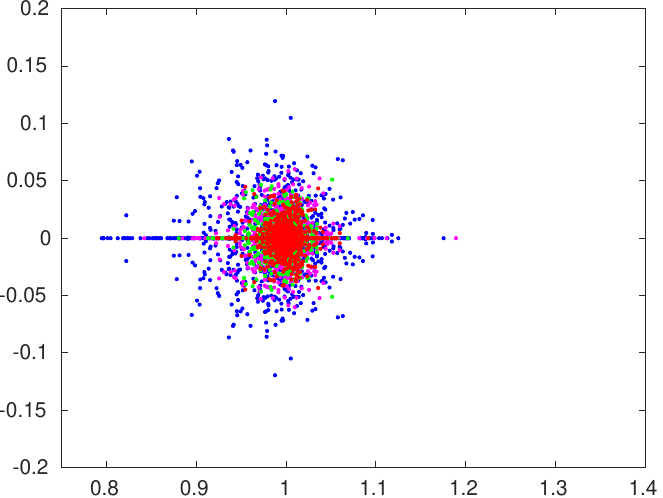}}\\
\subfloat[][Operators formed from fixed sparsity GMRES polynomials, relative drop tolerances applied to the resulting $\mat{R}$ and $\mat{P}$ and relative drop tolerances applied to the resulting $\mat{A}_\textrm{coarse}$.]{\label{fig:spectrum_sparse}\includegraphics[width =0.45\textwidth]{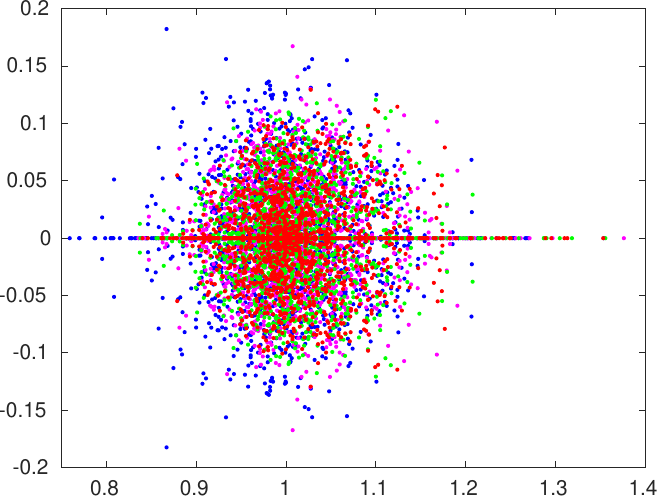}}
\caption{Eigenvalue distribution of $\mat{A}_\textrm{coarse}^{-1} \mat{S}$ on the second level of a pure streaming problem, where $\mat{S}$ is the exact Schur complement formed from the ideal restrictor and prolongator (i.e., the exact coarse-grid matrix) and $\mat{A}_\textrm{coarse}$ is the approximate coarse grid matrix formed from our approximate ideal restrictor and prolongator. The colours correspond to different GMRES polynomial orders for approximating $\mat{A}_\textrm{ff}^{-1}$, with \textcolor{blue}{$m=1$}, \textcolor{magenta}{$m=2$}, \textcolor{green}{$m=3$} and \textcolor{red}{$m=4$}. }
\label{fig:spectrum}
\end{figure}
As mentioned in \secref{sec:frozen}, the low-order polynomials we use and the fixed sparsity described in \secref{sec:fixed_sparsity} may impact the operators in our multigrid hierarchy. Given this we examine the impact of our approximations to $\mat{A}_\textrm{ff}^{-1}$ and the resulting operators by considering the spectrum of $\mat{A}_\textrm{coarse}^{-1} \mat{S}$ and $\mat{A}_\textrm{coarse}$, where $\mat{A}_\textrm{coarse}$ is the coarse matrix formed from our approximate operators and $\mat{S}$ is the exact coarse grid matrix.

As an example, we use AIRG on a pure streaming problem with an unstructured grid, with two levels of uniform refinement in angle (giving 16 angles). \fref{fig:spectrum} plots the the spectrum of $\mat{A}_\textrm{coarse}^{-1} \mat{S}$ and ideally it should approach one. \fref{fig:spectrum_exact} shows the result of constructing our coarse grid matrix with increasing orders of our GMRES polynomial, but without fixing the sparsity of the matrix powers, so we are using $q_0(\mat{A}_\textrm{ff})$, $q_1(\mat{A}_\textrm{ff})$, $q_2(\mat{A}_\textrm{ff})$ and $q_3(\mat{A}_\textrm{ff})$ exactly. We can see increasing the order increases the accuracy of our resulting coarse grid matrix, as would be expected, with the eigenvalues converging to one. The radius of a circle that bounds the eigenvalues reduces from 0.1760 to 0.0072 with $m=1$ to $m=4$, respectively.  

\fref{fig:spectrum_coarse_mat_not_sparse} shows the result of introducing both the fixed sparsity matrix powers discussed in \secref{sec:fixed_sparsity} and introducing relative drop tolerances to the resulting $\mat{R}$ and $\mat{P}$ operators. For the restrictor we drop any entry in a row that is less than 0.1 times the maximum absolute row entry, and for the prolongator we keep only the largest entry in each row. We can see these approximations are detrimental to $\mat{A}_\textrm{coarse}$ compared with \fref{fig:spectrum_exact}, with the eigenvalues further from one. The bounding circle at all orders has greater radius, with 0.2533 for $m=1$ and 0.0913 for $m=4$. 

Finally, \fref{fig:spectrum_sparse} shows the results from using the same fixed sparsity matrix powers and drop tolerances on $\mat{R}$ and $\mat{P}$ while also introducing a relative drop tolerance on the resulting $\mat{A}_\textrm{coarse}$, where we drop any entry in a row that is less than 0.1 times the maximum absolute row entry. We again see this further degrades our approximate coarse grid matrix, with the effect of increasing the order of our GMRES polynomials diminished; the bounding circle goes from 0.3539 to 0.3532 with $m=1$ to $m=4$, respectively.

\begin{figure}[th]
\centering
\subfloat[][The \textcolor{red}{$\times$} are eigenvalues of $\mat{A}_\textrm{coarse}$ with fixed sparsity GMRES polynomial, the \textcolor{red}{$\ast$} are with fixed sparsity GMRES polynomial and relative drop tolerances applied to the resulting $\mat{R}$ and $\mat{P}$.]{\label{fig:near_null_space_sparse_vs_not}\includegraphics[width =0.45\textwidth]{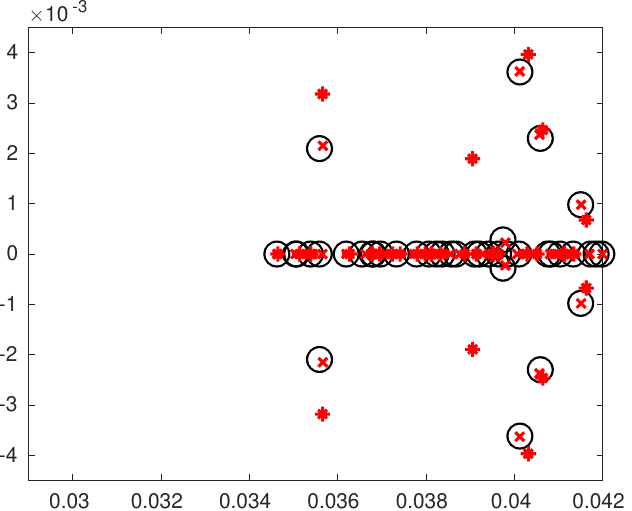}}\hspace{0.1cm}
\subfloat[][The \textcolor{red}{.} are eigenvalues of $\mat{A}_\textrm{coarse}$ with fixed sparsity GMRES polynomial, relative drop tolerances applied to the resulting $\mat{R}$ and $\mat{P}$ and relative drop tolerances applied to the resulting $\mat{A}_\textrm{coarse}$.]{\label{fig:near_null_space_sparse_coarse}\includegraphics[width =0.45\textwidth]{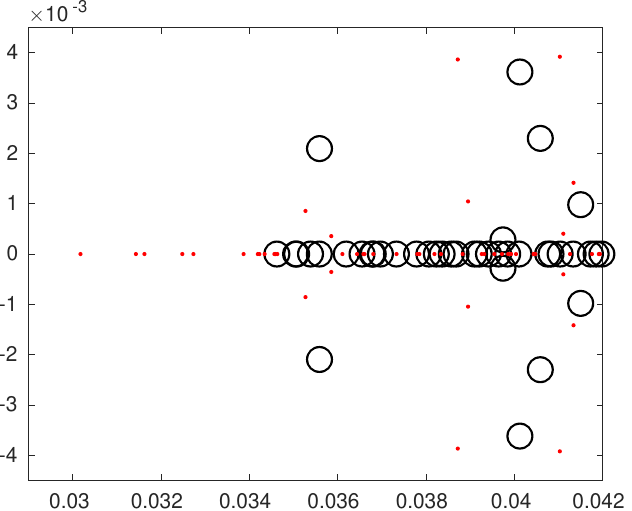}}
\caption{Eigenvalue distribution for the smallest eigenvalues of $\mat{A}_\textrm{coarse}$ and $\mat{S}$ on the second level of a pure streaming problem, where $\mat{S}$ is the exact Schur complement formed from the ideal restrictor and prolongator (i.e., the exact coarse-grid matrix) whose eigenvalues are denoted with the black ``o''. The red symbols are the eigenvalues of $\mat{A}_\textrm{coarse}$, which is the approximate coarse grid matrix formed from our approximate ideal restrictor and prolongator given a GMRES polynomial approximation of $\mat{A}_\textrm{ff}^{-1}$ with \textcolor{red}{$m=4$} and different sparsification applied to operators.}
\label{fig:near_null}
\end{figure}

It is clear that introducing additional sparsity into our GMRES polynomials and resulting operators degrades our coarse matrix. We examine this further by plotting the smallest eigenvalues of $\mat{A}_\textrm{coarse}$ and $\mat{S}$ in \fref{fig:near_null}. In the limit of ideal operators we know the near-nullspace vectors are preserved, but we would like to verify that with an approximate ideal restrictor and approximate ideal prolongator this is still the case. We can see that in \fref{fig:near_null_space_sparse_vs_not} that the GMRES polynomials with $m=4$ and fixed sparsity does an excellent job capturing the smallest eigenvalues. Furthermore introducing both fixed sparsity to the GMRES polynomial and drop tolerances on $\mat{R}$ and $\mat{P}$ results in reasonable approximations. We can see in \fref{fig:near_null_space_sparse_coarse} that introducing the drop tolerances on the resulting $\mat{A}_\textrm{coarse}$ results in small eigenvalues that do not match the exact coarse grid matrix.

These results indicate that our low-order GMRES polynomials with fixed sparsity and the introduction of drop tolerances to our approximate ideal $\mat{R}$ and $\mat{P}$ results in an excellent approximation for the coarse grid streaming operator. It is well known that introducing additional sparsity to multigrid operators can harm the resulting operators, and it is clear from these results that introducing a drop tolerance to our coarse grid matrix has the biggest impact. In a multilevel setting however, we find that doing this often result in the best performance, with the slight increase in iterations balanced by the reduced complexity; care must be taken to not make the drop tolerance too high.
\section{CF splitting}
\label{sec:F and C point selection}
All multigrid/multilevel methods require the formation of a hierarchy of ``grids''; LDU methods and reduction multigrids like in this work require the selection of a subset of DOFs defined as ``fine'' and ``coarse''. For asymmetric linear systems, CF splitting algorithms often result in coarse grids with directionality (i.e., they result in a semi-coarsening), typically through heuristic methods that identify strong connections in matrix entries (e.g, see \cite{Mavriplis1999}) with algorithms like CLJP, PMIS, HMIS, etc \cite{Henson2002}) or through compatible relaxation \cite{Brandt2000, Brannick2007, Brannick2010}. We would like the CF splitting to produce a well-conditioned $\mat{A}_\textrm{ff}$ on each level without giving a large grid or operator complexity across the hierarchy. The effectiveness of some of the approximations used in the literature (described in \secref{sec:airg}) for $\mat{A}_\textrm{ff}^{-1}$ also clearly depend on the sparsity of $\mat{A}_\textrm{ff}$ produced by a CF splitting.

Previous works have used various CF splittings, including those that produce a maximally-independent set, giving a diagonal $\mat{A}_\textrm{ff}$ that is easily inverted \cite{Saad1996}; or if a block-independent set is generated then $\mat{A}_\textrm{ff}$ is block-diagonal and the blocks can be inverted directly \cite{Saad1999, Saad2002}. With a more general CF splitting \cite{Saad2005} used ILU factorisations to approximate $\mat{A}_\textrm{ff}^{-1}$. \cite{MacLachlan2007} produced CF splittings specifically for reduction multigrids and LDU methods that are targeted at producing a diagonally dominant $\mat{A}_\textrm{ff}$.

In this work we use traditional CF splitting algorithms (like in \cite{Southworth2017}) as we find they perform well enough and parallel implementations are readily available. \secref{sec:Results} presents the results from using the lAIR implementation in \textit{hypre} and we found that using the Falgout-CLJP algorithm in \textit{hypre} resulted in good CF splittings. In order to make fair comparisons, we show results from using AIRG with the same algorithm.
\section{Work estimates}
\label{sec:Work estimates}
One of the key metrics we use to quantify the performance of the iterative methods tested is the number of Work Units (WUs) required to solve our linear systems; this is a FLOP count scaled by the number of FLOPs required to compute a matvec. We present several different WU calculations, each of which is scaled by a different matvec FLOP count. This is in an attempt to show fair comparisons against other multigrid methods, along with source iteration.

To begin, we must first establish a FLOP count for all the different components of our iterative methods. We begin with our definition of the Cycle Complexity (CC) of AIRG. The CC is the amount of work performed during a single V-cycle, scaled by the number of nnzs in the top-grid matrix. Our calculation of the CC includes the work performed during smoothing and grid-transfer operators; we use our definition of CC and WUs in all the results below. We define the work required to compute a matvec with our matrices on each level as $\{.\}^l$. For an assembled matrix we set this as the nnzs. This assumes fused-multiply-add (FMA) instructions are available and hence the cost of multiplying by $\mat{A}_\textrm{ff}$ for example is nnzs rather than $2\times$nnzs (this cost scales out of the CC anyway). If we consider a general linear system, $\mat{A} \mat{x} = \mat{b}$, the FLOP count for performing a single V-cycle with $l_{\textrm{max}}$ levels of AIRG is given by 
\begin{equation}
\textrm{FLOP}_\textrm{V}^\textrm{AIRG} = \{\hat{\mat{A}}^{-1}\}^{l_{\textrm{max}}} + \sum_{l=1}^{l=l_{\textrm{max}} - 1} v_\textrm{up} \{\hat{\mat{A}}_\textrm{ff}^{-1}\}^l + v_\textrm{up} \{\mat{A}_\textrm{ff}\}^l + \{\mat{A}_\textrm{fc}\}^l + \{\mat{R}\}^l + \{\mat{P}\}^l,
\label{eq:flop_v}
\end{equation}
where $v_\textrm{up}=2$ is the number of up F-point smooths and we perform one application of a GMRES polynomial approximation of $\hat{\mat{A}}^{-1}$ as a coarse grid solve on $l=l_\textrm{max}$.

In \secref{sec:Results}, we also show the results from using lAIR in \textit{hypre} with FCF-Jacobi smoothing; by default the CC output by \textit{hypre} doesn't include all the work associated with smoothing, residual calculation, etc, and hence we recompute it. Due to the use of FCF-Jacobi, the result of $\mat{A}_\textrm{fc} \mat{x}_\textrm{c}$ during the F smooths cannot be cached (similarly for the C-point smooths). We therefore compute
\begin{equation}
\textrm{FLOP}_\textrm{V}^\textrm{hypre} = \{\hat{\mat{A}}^{-1}\}^{l_{\textrm{max}}} + \sum_{l=1}^{l=l_{\textrm{max}} - 1} v_\textrm{up} (2 n_\textrm{F}^l + n_\textrm{C}^l) + v_\textrm{up} \left(2 \left(\{\mat{A}_\textrm{ff}\}^l + \{\mat{A}_\textrm{fc}\}^l \right) + \{\mat{A}_\textrm{cf}\}^l + \{\mat{A}_\textrm{cc}\}^l \right) + \{\mat{R}\}^l + \{\mat{P}\}^l,
\label{eq:flop_v_hypre}
\end{equation}
where $n_\textrm{F}^l$ and $n_\textrm{C}^l$ are the number of F and C-points on level $l$, respectively and given we only use one FCF-Jacobi in \textit{hypre} as the smoother on each level, $v_\textrm{up}=1$. The cycle complexity (for either \textit{hypre} or AIRG) is then given by
\begin{equation}
\textrm{CC} = \frac{\textrm{FLOP}_\textrm{V}}{\{\mat{A}\}^{1}}.
\label{eq:CC}
\end{equation}

Any matvec that involves scatter should be computed matrix-free and we denote that with an ``mf'' subscript. Given our sub-grid scale discretisation, we need to account for the cost of computing the source on the rhs of \eref{eq:SGS_sep_stream}, the fine-scale solution $\bm{\Theta}$ and the addition of the coarse and fine scale solutions to form $\bm{\psi}$. These are given by
\begin{equation}
\textrm{FLOP}_\textrm{source} = \{\mat{B}\}_\textrm{mf} + \{\hat{\mat{D}}^{-1}\}, \quad \textrm{FLOP}_\textrm{SGS} = \{\mat{C}\}_\textrm{mf} + \{\hat{\mat{D}}^{-1}\}, \quad \textrm{FLOP}_{\bm{\psi}} = \textrm{NDDOFs}
\label{eq:flop_source}
\end{equation}

The FLOP count of one iteration of our angular preconditioner \eref{eq:dsa} is given by 
\begin{equation}
\textrm{FLOP}_\textrm{angle} = 2 \times \textrm{NCDOFs} + 4.5 \times \{\mat{D}_\textrm{diff}\}.
\label{eq:flop_angle}
\end{equation}
We investigated using AIRG to invert the diffusion operator, but found it difficult to beat the default \textit{boomerAMG} implementation in \textit{hypre} (i.e., not lAIR), which is unsurprising given \textit{hypre} has been heavily optimised for such elliptic operators. The factor of 4.5 comes from the cycle complexity of running \textit{boomerAMG} on a heavily refined spatial grid and as might be expected we see the cycle complexity plateaus to around this value (hence we have an upper bound on work on less refined grids). 

We must now quantify the cost of our matrix-free matvecs. As mentioned in \secref{sec:sub-grid} there are numerous ways we could compute such a matvec, depending on how much memory we have available; \secref{sec:iterative} discussed that we store the streaming/removal operator, $\mat{M}_\Omega$ to precondition with and hence we use that in our matvec. The additional cost therefore comes with scattering (which we assume is isotropic in this work) and hence we have
\begin{multline}
\{\mat{A} - \mat{B}\hat{\mat{D}}^{-1} \mat{C}\}_\textrm{mf} = \{\mat{M}_\Omega\} + \sum_{i \in \textrm{cg nodes}} 2 \times \delta(i)\times \textrm{NCDOFs}(i) + \\ \sum_{i \in \textrm{dg nodes}} \delta(i) \times (2 \times \textrm{NDDOFs}(i) + 2 \times n_\textrm{nodes} + 1) + \sum_{e \in \textrm{eles}} 3 \times \delta(i) \times \{\hat{\mat{D}}^{-1}_e\}
\label{eq:mf_work}
\end{multline}
where $n_\textrm{nodes}$ is the number of spatial nodes on our DG elements (3 in 2D or 4 in 3D with tri/tets), $\{\hat{\mat{D}}^{-1}_e\}$ is the number of non-zeros in our stored block approximation on a given element (the factor of 3 comes from one application of $\hat{\mat{D}}^{-1}$ and one of $\mat{B}_\Omega$ and $\mat{C}_\Omega$ which have the same sparsity); this sparsity depends on the angles present on each DG node, but is at a maximum when uniform angle is used and for each angle present on all nodes of an element we have a $n_\textrm{nodes} \times n_\textrm{nodes}$ block. $\textrm{N*DOFS}(i)$ is the number of DOFs on an individual (CG or DG) spatial node $i$ and $\delta(i)$ is 1 or 0 depending on the presence of scatter; for a CG node this is zero if every element connected to node $i$ has a zero scatter cross-section, and one otherwise, for a DG node this is zero if the element containing node $i$ has a zero cross-section, one otherwise. The calculation in \eref{eq:mf_work} includes the work required to map to/from Legendre space on both our coarse (CG) and fine (DG) spatial meshes in order to apply the scatter component in $\mat{A}, \mat{B}$ and $\mat{C}$. Similar expressions are used for the individual $ \{\mat{B}\}_\textrm{mf}$ and $ \{\mat{C}\}_\textrm{mf}$ in \eref{eq:flop_source}.

We now have all the components required to calculate our WUs. We begin with a simple definition, where we use AIRG as a preconditioner on the assembled matrix, $\mat{A} - \mat{B}\hat{\mat{D}}^{-1} \mat{C}$, that could include scatter and is hence non-scalable. If $n_{\textrm{its}}$ is the number of outer GMRES iterations performed then the total FLOPs are
\begin{equation}
\textrm{FLOP}_\textrm{full} = n_{\textrm{its}} \left(\{\mat{A} - \mat{B}\hat{\mat{D}}^{-1} \mat{C}\} + \textrm{FLOP}_\textrm{V} \right) + \textrm{FLOP}_\textrm{source} + \textrm{FLOP}_\textrm{SGS} + \textrm{FLOP}_{\bm{\psi}},
\end{equation}
and hence the WUs 
\begin{equation}
\textrm{WUs}^\textrm{full} = \frac{\textrm{FLOP}_\textrm{full}}{\{\mat{A} - \mat{B}\hat{\mat{D}}^{-1} \mat{C}\}} .
\label{eq:WU_full}
\end{equation}
If instead we use the additively preconditioned iterative method defined in \secref{sec:iterative} along with our matrix-free matvec, our total FLOPs are
\begin{equation}
\textrm{FLOP}_\textrm{mf} = n_{\textrm{its}} \left(\{\mat{A} - \mat{B}\hat{\mat{D}}^{-1} \mat{C}\}_\textrm{mf} + \textrm{FLOP}_\textrm{V} + \textrm{FLOP}_\textrm{angle} \right) + \textrm{FLOP}_\textrm{source} + \textrm{FLOP}_\textrm{SGS} + \textrm{FLOP}_{\bm{\psi}},
\label{eq:flop_mf}
\end{equation}
We then scale these FLOP counts in several different ways. Firstly there is the WUs required to compute a matrix-free matvec of our sub-grid scale discretisation, namely
\begin{equation}
\textrm{WUs}^\textrm{mf} = \frac{\textrm{FLOP}_\textrm{mf}}{\{\mat{A} - \mat{B}\hat{\mat{D}}^{-1} \mat{C}\}_\textrm{mf}}.
\label{eq:WU_mf}
\end{equation}
In order to make rough comparisons with a traditional DG FEM source iteration method, we can scale by the work required to compute a matrix-free matvec with a DG FEM. In order to not unfairly disadvantage a DG discretisation, we assume the DG streaming operator is stored in memory, and hence the FLOPs required to compute a single DG matvec is
 \begin{equation}
\textrm{FLOP}_\textrm{DG} = \frac{5}{3}\{\hat{\mat{D}}^{-1}\} + \sum_{i \in \textrm{dg nodes}} \delta(i) \times (2 \times \textrm{NDDOFs}(i) + n_\textrm{nodes}).
\end{equation}
The factor of 5/3 comes from the jump terms (in 2D) that are not included in the nnzs of our sparsified DG matrix $\hat{\mat{D}}^{-1}$. The work units scaled by this quantity are therefore
\begin{equation}
\textrm{WUs}^\textrm{DG} = \frac{\textrm{FLOP}_\textrm{full}}{\textrm{FLOP}_\textrm{DG}} \quad \textrm{or} \quad \frac{\textrm{FLOP}_\textrm{mf}}{\textrm{FLOP}_\textrm{DG}}.
\label{eq:WU_dg}
\end{equation}
Again we note that with scattering we have $\{\mat{A} - \mat{B}\hat{\mat{D}}^{-1} \mat{C}\}_\textrm{mf} \approx 1.8 \times \textrm{FLOP}_\textrm{DG}$. 
\section{Results}
\label{sec:Results}
Outlined below are several examples problems in both the streaming and scattering limits, designed to test the performance of AIRG and our additively preconditioned iterative method. We solve our linear systems with GMRES(30) to a relative tolerance of 1\xten{-10}, with an absolute tolerance of 1\xten{-50} and use an initial guess of zero unless otherwise stated. We should note that we use AIRG (and lAIR) on our matrices without relying on any (potential) block structure, for example in the streaming limit with uniform angle we could use our multigrid on each of the angle blocks separately. We also don't scale our matrices; for example we could view a diagonal scaling as preconditioning the outer GMRES iteration and/or the GMRES polynomials in our multigrid, but we did not find it necessary. 

When using AIRG we perform zero down smooths and two up F-point smooths (our C-points remain unchanged during smoothing). On the bottom level of our multigrid we use one Richardson iteration to apply a GMRES polynomial approximation of the coarse matrix. Unless otherwise noted we use 3rd order ($m=4$) GMRES polynomials as described in \secref{sec:fixed_sparsity} on all levels including the bottom; we compare using the polynomials with and without fixed sparsity. We use a 1-point ideal prolongator which computes $\mat{W}$ and then keeps the biggest absolute entry per row. We only use isotropic scatter in this work. For both AIRG and lAIR, we use the row-wise infinity norm to define any drop tolerances. When using the lAIR implementation in \textit{hypre}, we use zero down smooths and one iteration of FCF-Jacobi for up smooths, while on the bottom level we use a direct-solve, as we found these options resulted in the lowest cost/best scaling. We searched the parameter space to try and find the best values for drop tolerances, strength of connections, etc when using lAIR and AIRG; these searches were not exhaustive however, and it is possible there are more optimal values.

All timing results are taken from compiling our code, PETSc 3.15 and \textit{hypre}  with ``-O3'' optimisation flag. We compare timing results between our PETSc implementation of AIRG and the \textit{hypre} implementation of lAIR. As such we try and limit the impact of different implementation details. Given this, when calculating the setup time we exclude the CF splitting time, the time required to drop entries from matrices (as the PETSc interfaces require us to take copies of matrices), and to extract the submatrices $\mat{A}_\textrm{ff}$, $\mat{A}_\textrm{fc}$, etc. These should all be relatively low cost parts of the setup, are shared by both AIRG and lAIR and should scale with the nnzs. The setup time we compare is therefore that required to form the restrictors, prolongators and coarse matrices. Also we should note that with AIRG our setup time is an upper bound, as we have not built an optimised matmatmult for building our fixed sparsity GMRES polynomials (we know the sparsity of the matrix powers is the same as the two input matrices). As such we compute a standard matmatmult in PETSc and then drop entries (although we do provide a flop count for a matmatmult with fixed sparsity). When timing lAIR, we use the PETSc interface to \textit{hypre} and hence we run two solves (and set the initial condition to zero in both). The \textit{hypre} setup occurs on the 0th iteration of the first solve, so a second solve allows us to correctly measure just the solve time. We only show solve times for problems with pure streaming, as our implementation of the P$^0$ matrix-free matvec with scattering is not well optimised.

All tabulations of memory used are scaled by the total NDOFs in $\bm{\psi}$ in \eref{eq:SGS_full}, i.e., NDOFs=NCDOFs + NDDOFs. Included in this figure is the memory required to store the GMRES space, both additive preconditioners (if required) and hence the AIRG hierarchy, $\hat{\mat{A}}_\textrm{ff}^{-1}$, separate copies of $\mat{A}_\textrm{ff}, \mat{A}_\textrm{fc}, \mat{A}_\textrm{cf}$ and $\mat{A}_\textrm{cc}$, the diffusion operator and temporary storage.  We do not report the memory use of \textit{hypre}, although given the operator complexities it is similar to AIRG. 
\subsection{AIRG on the matrix $\mat{A} - \mat{B}\hat{\mat{D}}^{-1} \mat{C}$}
\label{sec:air_full_mat}
In this section, we build the matrix $\mat{A} - \mat{B}\hat{\mat{D}}^{-1} \mat{C}$ and use this matrix as a preconditioner, applied with 1 V-cycle of either AIRG or lAIR. As such we do not use the iterative method described in \secref{sec:iterative}, nor do we use the matrix-free matvec described in \secref{sec:sub-grid}. As mentioned using an iterative method that relies on the full operator is not practical with scattering given the nonlinear increase in nnzs with angular refinement, but we wish to show our methods are still convergent in the scattering (diffuse) limit. Instead, \secref{sec:mf_iterative} shows the results from using the iterative method in \secref{sec:iterative}. 

Our test problem is a $3 \times 3$ box with a source of strength 1 and size $0.2 \times 0.2$ in the centre of the domain. We apply vacuum conditions on the boundaries and discretise this problem with unstructured triangles and ensure that our grids are not semi-structured (e.g., we don't refine coarse grids by splitting elements). We use uniform level 1 refinement in angle, with 1 angle per octant (similar to S$_2$).
\subsubsection{Pure streaming problem}
\label{sec:pure_stream}
For the pure streaming problem we set the total and scatter cross-sections to zero. To begin, we examine the performance of AIRG and lAIR with spatial refinement and a fixed uniform level 1 angular discretisation (i.e, with 4 angles in 2D). \tref{tab:2D_stream_hypre} shows that using distance 1 lAIR in this problem, with Falgout-CLJP CF splitting results in growth in both the iteration count and work. We could not find a combination of parameters that results in scalability with lAIR; increasing the number of FCF smooths to 3 results in an iteration count with similar growth, namely 15, 14, 14, 17, 19 and 22, but with a cycle complexity at the finest spatial refinement of 11.2 and hence 271 WUs. Even using distance 2 lAIR didn't result in scalability, as shown in \tref{tab:2D_stream_hypre_disttwo} where we can see a slightly decreased iteration count, with higher cycle and operator complexities, resulting in a similar number of WUs. Using both distance 2 lAIR and 3 FCF smooths still results in growth, with iteration counts of 15, 13, 14, 16, 17 and 19, with a cycle complexity of 15.9 and hence 323 WUs at the finest spatial refinement. 
\begin{table}[ht]
\centering
\begin{tabular}{ c c | c c c c c c}
\toprule
CG nodes & NDOFs & $n_\textrm{its}$ & CC & Op Complx &  WUs$^\textrm{full}$ & WUs$^\textrm{DG}$ & Memory\\
\midrule
97 & 2.4\xten{3} & 26 & 2.96 & 1.5 & 106 & 26.3 & -\\
591 & 1.6\xten{4} & 25 & 3.5 & 1.77 & 116 & 27.8 & -\\
2313	 & 6.3\xten{4} & 28 & 3.8 & 1.9 & 138 & 32.6 & - \\
9166	 & 2.5\xten{5} & 31 & 4.1 & 2.0 & 159 & 37.4 & - \\
35784 & 9.9\xten{5} & 36 & 4.2 & 2.1 & 189 & 44.5 & - \\
150063 & 4.2\xten{6} & 42 & 4.3 & 2.16 & 225 & 52.6 & - \\
\bottomrule  
\end{tabular}
\caption{Results from using distance 1 lAIR in \textit{hypre} on a pure streaming problem in 2D with CF splitting by the \textit{hypre} implementation of Falgout-CLJP with a strong threshold of 0.2, drop tolerance on $\mat{A}$ of 0.0075 and $\mat{R}$ of 0.025 and a strong $\mat{R}$ threshold of 0.25.}
\label{tab:2D_stream_hypre}
\end{table}
\begin{table}[ht]
\centering
\begin{tabular}{ c c | c c c c c c}
\toprule
CG nodes & NDOFs & $n_\textrm{its}$ & CC & Op Complx &  WUs$^\textrm{full}$ & WUs$^\textrm{DG}$ & Memory\\
\midrule
97 & 2.4\xten{3} & 26 & 3.2 & 1.58 & 112 & 27.9  & -\\
591 & 1.6\xten{4} & 24 & 3.7 & 1.85 & 116 & 28 & -\\
2313	 & 6.3\xten{4} & 27 & 4.1 & 2.04 & 141 & 33.5 & - \\
9166	 & 2.5\xten{5} & 30 & 4.4 & 2.16 & 164 & 38.6 & - \\
35784 & 9.9\xten{5} & 34 & 4.6 & 2.26 & 192 & 44.9 & - \\
150063 & 4.2\xten{6} & 38 & 4.7 & 2.32 & 219 & 51.1 & - \\
\bottomrule  
\end{tabular}
\caption{Results from using distance 2 lAIR in \textit{hypre} on a pure streaming problem in 2D with CF splitting by the \textit{hypre} implementation of Falgout-CLJP with a strong threshold of 0.2, drop tolerance on $\mat{A}$ of 0.0075 and $\mat{R}$ of 0.025 and a strong $\mat{R}$ threshold 0.25.}
\label{tab:2D_stream_hypre_disttwo}
\end{table}

We also tried decreasing the strong $\mat{R}$ threshold to 1 \xten{-7} in case some neighbours were being excluded, but this resulted in very little change in the iteration count, while the number of nnzs in the restrictor (and hence the setup time) grew considerably. Preliminary investigation suggests we would need to include neighbours at greater distance than two (along with only F-point smoothing, rather than FCF), but this increases the setup cost considerably. We also note that using nAIR (at several different orders) in this case results in a similar iteration count (even with diagonal scaling of our operators); this is likely due to the lack of lower triangular structure in our discretisation. 
\begin{table}[ht]
\centering
\begin{tabular}{ c c | c c c c c c c c}
\toprule
CG nodes & NDOFs & $n_\textrm{its}$ & CC & Op. Complx & WUs$^\textrm{full}$ & WUs$^\textrm{DG}$ & Memory\\
\midrule
97 & 2.4\xten{3} & 11 & 5.58 & 1.96 & 83 & 20.4 & 11.9\\
591 & 1.6\xten{4} & 10 & 5.85 & 2.48 & 79 & 18.9 & 11.7\\
2313	 & 6.3\xten{4} & 8 & 6.4 & 2.88 & 70 & 16.6 & 12.2 \\
9166	 & 2.5\xten{5} & 8 & 6.7 & 3.16 & 73 & 17.1 & 12.4 \\
35784 & 9.9\xten{5} & 9 & 6.9 & 3.36 & 82 & 19.3 & 12.5 \\
150063 & 4.2\xten{6} & 9 & 7.07 & 3.48 & 84 & 19.6 & 12.7 \\
\bottomrule  
\end{tabular}
\caption{Results from using AIRG with $m=4$ and without fixed sparsity on a pure streaming problem in 2D with CF splitting by the \textit{hypre} implementation of Falgout-CLJP with a strong threshold of 0.2, drop tolerance on $\mat{A}$ of 0.0075 and $\mat{R}$ of 0.025.}
\label{tab:2D_stream_hypre_fc_unconstrained}
\end{table}

Tables \ref{tab:2D_stream_hypre_fc_unconstrained} \& \ref{tab:2D_stream_hypre_fc} however shows that using AIRG with a third-order GMRES polynomial and Falgout-CLJP CF splitting results in less work with smaller growth. Using GMRES polynomials without fixed sparsity requires 84 WUs at the highest level of refinement, compared to 75 with fixed sparsity. The work in \tref{tab:2D_stream_hypre_fc_unconstrained} has plateaued, however the work with fixed sparsity is growing slightly with spatial refinement. Compared to lAIR, we see that AIRG with sparsity control needs three times less work to solve our pure streaming problem. Rather than use our one-point ideal prolongator, we also investigated using $\mat{W}$ with the same drop tolerances as applied to $\mat{Z}$. With fixed-sparsity this resulted in 11, 10, 9, 9, 9 and 10 iterations, with 67, 68, 66, 68, 70, and 78 WUs. The slight increase in cycle complexity is compensated by using one fewer iterations at the finest level, but overall we see similar work. In an attempt to ascertain the effect of using our one-point approximation to the ideal prolongator, we also constructed a classical one-point prolongator as in \cite{Southworth2017, Manteuffel2019} where each F-point is injected from its strongest C-point neighbour. With fixed-sparsity this results in the same number of iterations and WUs as in \tref{tab:2D_stream_hypre_fc}, which confirms that the classical prolongator is not responsible for the poor performance of lAIR. The benefit of using a classical one-point prolongator in this fashion is we require one fewer matmatmults on each level of our setup. For the rest of this paper we use the one-point approximation to the ideal operator, but it is clear there is a range of practical operators (with different sparsification strategies) we could use with AIRG. 

\tref{tab:2D_stream_hypre_fc} shows we can solve our streaming problem with the equivalent of approximately 18 DG matvecs. We can also see that the plateau in the operator complexity results in almost constant memory use, at 10 copies of the angular flux. We should also note that we can use AIRG as a solver, rather than as a preconditioner. We see the same iteration count in pure streaming problems and the lack of the GMRES space means we only need memory equivalent to approx. 5 copies of the angular flux; this is the amount of memory that would be required to store just a DG streaming operator in 2D. This shows that our discretisation and iterative method are low-memory in streaming problems.
\begin{table}[ht]
\centering
\begin{tabular}{ c c | c c c c c c c c}
\toprule
CG nodes & NDOFs & $n_\textrm{its}$ & CC & Op. Complx & WUs$^\textrm{full}$ & WUs$^\textrm{DG}$ & Memory\\
\midrule
97 & 2.4\xten{3} & 12 & 3.7 & 1.97 & 67 & 16.5 & 10\\
591 & 1.6\xten{4} & 10 & 4.1 & 2.5 & 62 & 14.7 & 10\\
2313	 & 6.3\xten{4} & 9 & 4.4 & 2.87 & 60 & 14.1 & 10.2 \\
9166	 & 2.5\xten{5} & 10 & 4.6 & 3.17 & 67 & 15.8 & 10.3 \\
35784 & 9.9\xten{5} & 10 & 4.74 & 3.36 & 69 & 16 & 10.4 \\
150063 & 4.2\xten{6} & 11 & 4.84 & 3.5 & 75 & 17.6 & 10.4 \\
\bottomrule  
\end{tabular}
\caption{Results from using AIRG with $m=4$ and fixed sparsity on a pure streaming problem in 2D with CF splitting by the \textit{hypre} implementation of Falgout-CLJP with a strong threshold of 0.2, drop tolerance on $\mat{A}$ of 0.0075 and $\mat{R}$ of 0.025.}
\label{tab:2D_stream_hypre_fc}
\end{table}

\begin{figure}[th]
\centering
\subfloat[][The \textcolor{matlabblue}{$\times$} is for AIRG with fixed sparsity, \textcolor{foliagegreen}{$\times$} is for AIRG without fixed sparsity and \textcolor{deludedorange}{$\otimes$} is for lAIR.]{\label{fig:flop_build_R_lair_airg}\includegraphics[width =0.45\textwidth]{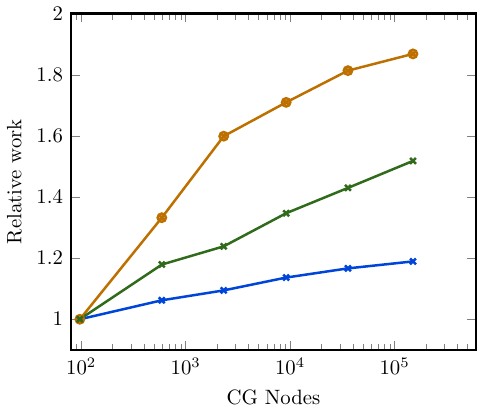}}
\caption{Sum of FLOPs required across all levels to compute $\mat{Z}$, scaled to the NDOFs, and then relative to the work required on the least refined spatial grid, for AIRG with $m=4$ (see \tref{tab:2D_stream_hypre_fc}) and distance 1 lAIR (see \tref{tab:2D_stream_hypre}) with Falgout-CLJP in a 2D pure streaming problem. The relative scaling is done separately for each line; they do not all cost the same at the coarsest resolution, we provide timings in Section \ref{sec:pure_stream}.}
\label{fig:setup_z}
\end{figure}

The cost of a solve must also be balanced by the cost of the setup of our multigrid. Our setup involves computing our GMRES polynomial approximations, followed by standard AMG operations, namely computing matrix-matrix products to form our restrictors/prolongators and coarse matrices on each level. \fref{fig:setup_z} begins by showing the relative amount of work required to compute $\mat{Z}$ for AIRG and distance 1 lAIR. For AIRG this is the sum of FLOPs required to compute $\hat{\mat{A}}_\textrm{ff}^{-1}$ and those to compute $-\mat{A}_\textrm{cf} \hat{\mat{A}}_\textrm{ff}^{-1}$. For lAIR, it is more difficult to calculate a FLOP count for the small dense solves which locally enforce $\mat{R}\mat{A}=\bm{0}$, as it is dependent on the BLAS implementation; instead we take the size of each of the $n \times n$ dense systems and compute the sum of $n^3$. As such we do not try and compare an absolute measure of the work required by both (we have timing results below). \fref{fig:setup_z} instead shows that the growth in work for both AIRG with fixed sparsity and lAIR begins to plateau with grid refinement; AIRG without fixed sparsity however grows with refinement, showing the necessity of the sparsity control on the matrix powers. 

\begin{figure}[th]
\centering
\subfloat[][The \textcolor{matlabblue}{$\times$} is the cost of computing $\hat{\mat{A}}_\textrm{ff}^{-1}$ for AIRG with fixed sparsity, the \textcolor{gaylordpurple}{$\times$} is with fixed sparsity computed with a standard matmatmult followed by dropping entries and the \textcolor{foliagegreen}{$\times$} is without fixed sparsity]{\label{fig:flop_build_gmres_flop}\includegraphics[width =0.45\textwidth]{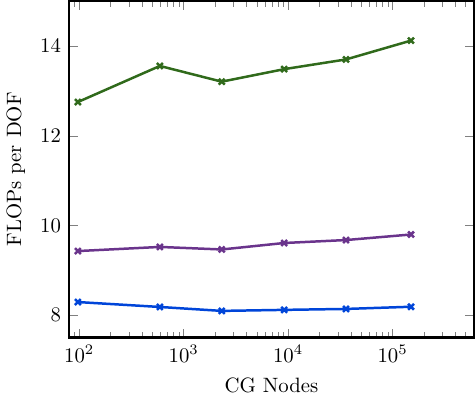}} \quad
\subfloat[][The dashed \textcolor{matlabblue}{$\times$} is the cost of the setup for AIRG with fixed sparsity, the \textcolor{foliagegreen}{$\times$} is without fixed sparsity, the \textcolor{matlabblue}{$\times$} is the setup plus solve with fixed sparsity, the \textcolor{foliagegreen}{$\times$} is without fixed sparsity. ]{\label{fig:flop_setup_plus_solve}\includegraphics[width =0.45\textwidth]{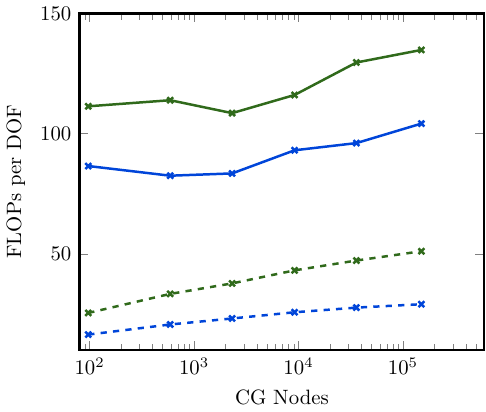}} \qquad
\caption{Sum of FLOPs required across all levels during the setup and solve, scaled to the NDOFs, for AIRG with $m=4$ and Falgout-CLJP in a 2D pure streaming problem (see Tables \ref{tab:2D_stream_hypre_fc} and \ref{tab:2D_stream_hypre_fc_unconstrained}).}
\label{fig:setup work}
\end{figure}

\fref{fig:flop_build_gmres_flop} shows that the cost of computing our GMRES polynomial approximations to $\hat{\mat{A}}_\textrm{ff}^{-1}$ with fixed sparsity is constant, at around 8 FLOPs per DOF. The plateauing growth seen in \fref{fig:setup_z} therefore comes from the matmatmult required to compute $-\mat{A}_\textrm{cf} \hat{\mat{A}}_\textrm{ff}^{-1}$. The FLOP count with fixed sparsity relies on a custom matmatmult algorithm that assumes the same sparsity in each component of the matmatmult, as in \eref{eq:sparsity_powers}. Computing the matrix powers with a standard matmatmult and then dropping any fill-in results in an almost constant number of FLOPs per DOF, at around 9.6, which is convenient as this makes it less necessary to build a custom matmatmult implementation. If we do not fix the sparsity of our polynomial approximations we can see growth in the cost, due to the increasingly dense matrix powers, as might be expected. 

\fref{fig:flop_setup_plus_solve} shows that the cost of our setup with fixed sparsity plateaus, and the total cost of our setup plus solve grows, due to the increase in work during the solve shown in \tref{tab:2D_stream_hypre_fc}, with 29\% growth from lowest to highest refinement. Without fixed sparsity we see less growth in \tref{tab:2D_stream_hypre_fc_unconstrained} from the solve, but higher growth in the setup. This results in similar growth in the total work, but the fixed sparsity case uses approximately 30\% fewer FLOPs. 

\begin{figure}[th]
\centering
\subfloat[][Solve time]{\label{fig:solve_times}\includegraphics[width =0.45\textwidth]{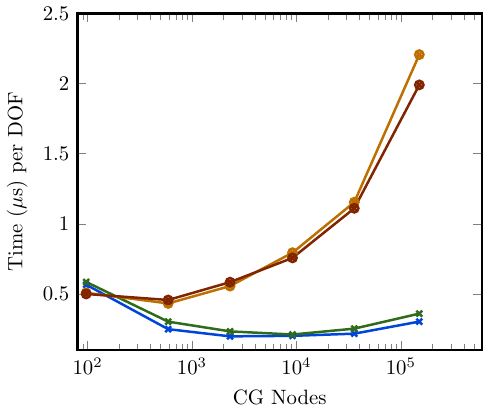}} \quad
\subfloat[][Dashed are time to compute $\hat{\mat{A}}_\textrm{ff}^{-1}$ for AIRG, solid are time to compute $\mat{Z}$]{\label{fig:z_times}\includegraphics[width =0.45\textwidth]{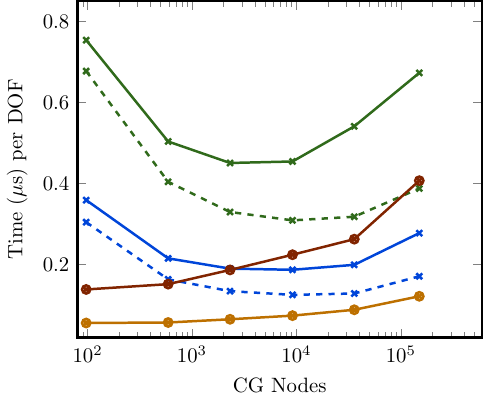}}\\
\subfloat[][Setup time]{\label{fig:setup_times}\includegraphics[width =0.45\textwidth]{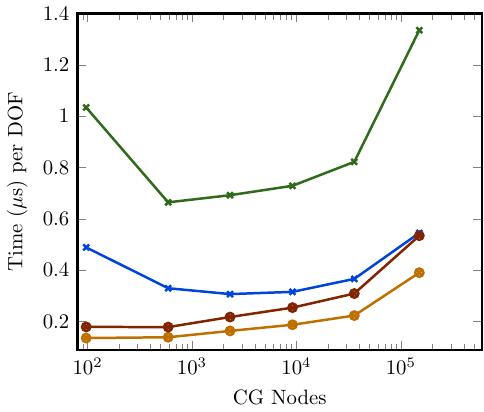}}\quad
\subfloat[][Total time]{\label{fig:total_times}\includegraphics[width =0.45\textwidth]{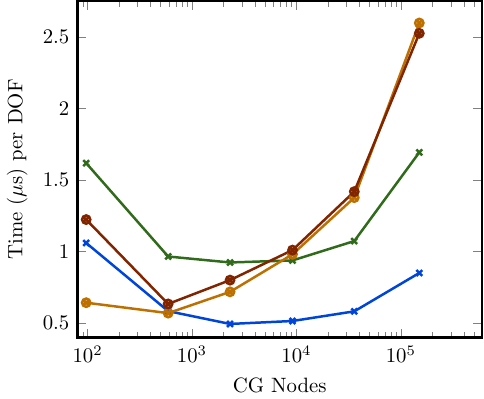}}
\caption{Timings per DOF for AIRG with $m=4$ and lAIR in a 2D pure streaming problem. The \textcolor{matlabblue}{$\times$} is AIRG with fixed sparsity with Falgout-CLJP, the \textcolor{foliagegreen}{$\times$} is AIRG without fixed sparsity and Falgout-CLJP, the \textcolor{deludedorange}{$\otimes$} is distance 1 lAIR with Falgout-CLJP, the \textcolor{fireenginered}{$\otimes$} is distance 2 lAIR with Falgout-CLJP.}
\label{fig:times}
\end{figure}

\fref{fig:times} shows results for the solve, $\mat{Z}$, setup and total time taken to solve our system with AIRG and lAIR. We can see in \fref{fig:solve_times} that the results match those in Tables 1-5, with AIRG with Falgout-CLJP coarsening and fixed sparsity giving the lowest solve times. We found a difference in the efficiency of our PETSc implementation vs the \textit{hypre} implementations however. If we modify the parameters used with AIRG and increase the work required to solve to roughly match that distance 1 lAIR in this problem, we found that there was a factor of almost 2 difference in the solve times. 

\fref{fig:z_times} shows the total cost of computing the approximate ideal restrictor, $\mat{Z}$. For AIRG we also show the time for computing the GMRES polynomial approximation to $\mat{A}_\textrm{ff}^{-1}$ separately. We can see for AIRG with fixed sparsity there is slight growth in the time to compute our polynomials, with much greater growth without fixed sparsity. Given the FLOP count in \fref{fig:flop_build_gmres_flop} we would expect the time with fixed sparsity to be constant. The further growth we see in the time to compute $\mat{Z}$ comes from the matmatmult to compute $-\mat{A}_\textrm{cf} \mat{A}_\textrm{ff}^{-1}$, which \fref{fig:flop_setup_plus_solve} suggests should plateau. We also see growth in the time taken to compute $\mat{Z}$ with lAIR, with higher growth for distance 2. Again \fref{fig:setup_z} shows the work estimate plateauing with spatial refinement. At the higher spatial refinements, the $\mat{Z}$ with AIRG, Falgout-CLJP coarsening and fixed sparsity costs more to compute than distance 1 lAIR, but less than distance 2. An AIRG implementation that takes advantage of the shared sparsity of the matrices when forming $\hat{\mat{A}}_\textrm{ff}^{-1}$ could reduce this setup further, with a reduction in the FLOPs required (by approx. 20\% as shown in \fref{fig:flop_build_gmres_flop}), and also in the cost of any symbolic computation. Given the substantially improved convergence shown in \tref{tab:2D_stream_hypre_fc} when compared to distance 2 lAIR in \tref{tab:2D_stream_hypre_disttwo}, this shows AIRG is an effective and relatively cheap way to compute approximate ideal operators in convection problems.

\fref{fig:setup_times} shows the setup times (which include the times to compute $\mat{Z}$ from \fref{fig:z_times}) and we can see that lAIR has the cheapest overall setup, with fixed sparsity AIRG coming in the middle, and AIRG without fixed sparsity giving the highest setup times. We expect AIRG without fixed sparsity to be increasing given the increased cost of computing the GMRES polynomial with spatial refinement, as shown in \fref{fig:flop_build_gmres_flop}, but we see increased setup time with AIRG with fixed sparsity and distance 1 lAIR, whereas the work estimates in Figures \ref{fig:setup_z} and \ref{fig:flop_setup_plus_solve} again suggests the setup times should plateau. We found that the time taken to compute the matmatmults used to build the transfer operators and coarse matrices is increasing more than the FLOP count would imply. Of course a FLOP count is not necessarily perfectly indicative of the time required in a matmatmult given the symbolic compute and memory accesses. It is possible more efficient matmatmult implementations are available. 

\fref{fig:total_times} shows the total time taken to setup and solve with our multigrid. Two of our AIRG results, namely using GMRES polynomials without and without fixed sparsity manage to beat lAIR. This helps show the promise of GMRES polynomials as part of an AIR-style multigrid; even without fixed sparsity they can be competitive. The fixed sparsity AIRG however results in a considerably decrease in total time, taking approx. 3$\times$ less time than distance 1 lAIR. Even if we try and remove the effect of implementation differences between lAIR and AIRG discussed above, by equating the solve time of lAIR with an AIRG result that requires similar WUs, fixed sparsity AIRG still has a total time of 0.65 $\times$ that of distance 1 lAIR. 

We also examined using SAIs to build an approximation of $\mat{A}_\textrm{ff}^{-1}$ with the fixed sparsity of $\mat{A}_\textrm{ff}$, as part of a reduction multigrid, like in \cite{Chow2003, Zaman2022}. We found approximations of $\mat{A}_\textrm{ff}^{-1}$ computed with SAIs comparable to our fixed sparsity GMRES polynomials with $m=4$. With spatial refinement, using SAIs instead of our GMRES polynomials as part of a reduction multigrid (with the same CF splitting, drop tolerances, etc as used in AIRG) gave 11, 10, 9, 10, 10 and 11 iterations and hence the same work units and solve times as AIRG given the matching fixed sparsity (except for the first spatial refinement). However we found it was more expensive to setup SAIs on each level when compared to our GMRES polynomials, by approximately 2$\times$; we used the \textit{ParaSails} implementation in \textit{hypre} for comparisons. Forming the fixed sparsity GMRES polynomials requires fewer FLOPs as the nnzs in $\mat{A}_\textrm{ff}$ grow compared to fixed sparsity SAI. If $n_\textrm{F}$ is the number of F-points, we require $m$ matvecs which scale linearly with the nnzs and a single QR factorisation of size $n_\textrm{F} \times (m+1)$, which doesn't depend on the nnzs. If $m>2$, we must also compute $m-2$ fixed sparsity matrix powers at cost $(m-2) s_\textrm{r} s_\textrm{c} n_\textrm{F}$, where $s_\textrm{r}$ and $s_\textrm{c}$ are the average number of nnzs per row and column of $\mat{A}_\textrm{ff}$, respectively. If we assume $s = s_\textrm{r} \approx s_\textrm{c}$ then the cost of computing the matrix-powers can be written as $s (m-2) \times \textrm{nnzs}(\mat{A}_\textrm{ff})$. In comparison, the SAI algorithm with the fixed sparsity of $\mat{A}_\textrm{ff}$ requires solving $n_\textrm{F}$ (local) least-squares problems. If we just consider the required $n_\textrm{F}$ QR factorisations, typically they would be of size $t \times s$, where $t$ is the number of entries in the ``shadow'' of a column with $s$ entries (typically $t > s$). We simplify here to the square $s \times s$, which is equivalent to considering a cheaper ``incomplete'' SAI \cite{Anzt2018a} and is similar to the compute required by lAIR. This requires a FLOP count of $s^2 \times \textrm{nnzs}(\mat{A}_\textrm{ff})$ (we have dropped the constants). For many problems a low polynomial order is acceptable and hence $(m-2) \ll s$; this is particularly true on lower multigrid levels where the average row or column sparsity (and hence the size of the shadow for a given row/column) can grow. There may be a scale, however, at which the extra local cost of setting up SAIs is balanced by the all-reduce required by our fixed sparsity GMRES polynomials. One of the benefits of using our GMRES polynomials however is that we can decrease the amount of communication required by decreasing the polynomial order below 2. 

\begin{figure}[th]
\centering
\subfloat[][Setup time]{\label{fig:setup_times_gmres}\includegraphics[width =0.45\textwidth]{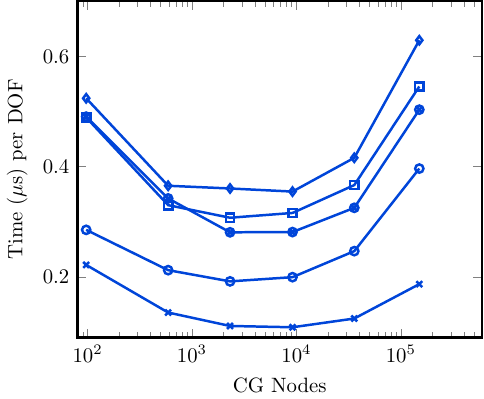}}\quad
\subfloat[][Total time]{\label{fig:total_times_gmres}\includegraphics[width =0.45\textwidth]{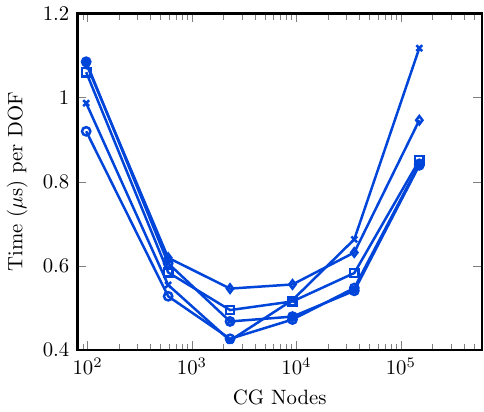}}
\caption{Timings per DOF for AIRG with fixed sparsity, Falgout-CLJP and with varying GMRES polynomial order in a 2D pure streaming problem. The \textcolor{matlabblue}{$\times$} is AIRG with $m=1$, \textcolor{matlabblue}{o} is $m=2$, \textcolor{matlabblue}{$\otimes$} is $m=3$, \textcolor{matlabblue}{$\square$} is $m=4$, \textcolor{matlabblue}{$\diamond$} is $m=5$.}
\label{fig:times_gmres}
\end{figure}

Given this, we examine the role of changing the GMRES polynomial order with AIRG and fixed sparsity, from zero to four ($m=1$ to $m=5$; the zeroth and first order polynomials implicitly have fixed sparsity) in \fref{fig:times_gmres}. We can see the zeroth order polynomial is very cheap to construct, but results in the highest total time. This is because the iteration count is higher. The increasingly higher order polynomials take longer to setup, although the difference between successive orders decreases, due to the shared sparsity. With increasing polynomial order, on the most refined spatial grid we have 40, 16, 12, 11 and 11 iterations to solve, with cycle complexities of 3.25, 4.67, 4.81, 4.84 and 4.85, respectively. We can see that all the polynomials between first and third order result in similar total times; the decreased cost of setup for the lower orders is balanced by the increase in iterations. We see however very similar growth in iteration count with spatial refinement with the first through fourth order polynomials; for example with first order GMRES polynomials ($m=2$) in \fref{fig:times_gmres} we have 15, 14, 12, 14, 14 and 16 iterations. This gives a higher overall amount of work than with $m=4$ (up to approximately 100 WUs at the highest refinement), but given the similar growth in iteration count and reduced communication in parallel during the setup, requiring only two matvecs and a single all-reduce, this may be a good choice in parallel.

Given the results with spatial refinement above, we chose to examine the role of angular refinement with AIRG with fixed sparsity and $m=4$. \tref{tab:2D_stream_angle} shows that with 3 levels of angular refinement on the third refined spatial grid, the iteration count increases slightly from 9 to 11, but the cycle and operator complexity are almost identical and hence we have fixed memory consumption. We do not show the timings as they scale as would be expected. \tref{tab:2D_stream_angle_hypre} shows that distance 1 lAIR requires a fixed amount of work with angular refinement, but this is roughly twice that of AIRG. 

\begin{table}[ht]
\centering
\begin{tabular}{ c c c | c c c c c c c}
\toprule
CG nodes & Angle lvl. & NDOFs & $n_\textrm{its}$ & CC & Op. Complx & WUs$^\textrm{full}$ & WUs$^\textrm{DG}$ & Memory\\
\midrule
2313 & 1 & 6.3\xten{4} & 9 & 4.4 & 2.87 & 60 & 14.1 & 10.2 \\
2313 & 2 & 2.5\xten{5} & 10 & 4.4 & 2.88 & 65 & 15.4 & 10.4 \\
2313 & 3 & 1\xten{6} & 11 & 4.4 & 2.87 & 70 & 16.7 & 10.4 \\
\bottomrule  
\end{tabular}
\caption{Results from using AIRG with $m=4$ and fixed sparsity on a pure streaming problem in 2D with CF splitting by the \textit{hypre} implementation of Falgout-CLJP with a strong threshold of 0.2, drop tolerance on $\mat{A}$ of 0.0075 and $\mat{R}$ of 0.025 with different levels of angular refinement.}
\label{tab:2D_stream_angle}
\end{table}
\begin{table}[ht]
\centering
\begin{tabular}{ c c c | c c c c c c c}
\toprule
CG nodes & Angle lvl. & NDOFs & $n_\textrm{its}$ & CC & Op Complx &  WUs$^\textrm{full}$ & WUs$^\textrm{DG}$ & Memory\\
\midrule
2313 & 1 & 6.3\xten{4} & 28 & 3.8 & 1.9 & 138 & 32.6 & - \\
2313 & 2 & 2.5\xten{5} & 27 & 3.7 & 2.36 & 133 & 31.7 & - \\
2313 & 3 & 1\xten{6} & 28 & 3.8 & 2.35 & 133 & 31.7 & - \\
\bottomrule  
\end{tabular}
\caption{Results from using distance 1 lAIR in \textit{hypre} on a pure streaming problem in 2D with CF splitting by the \textit{hypre} implementation of Falgout-CLJP with a strong threshold of 0.2, drop tolerance on $\mat{A}$ of 0.0075 and $\mat{R}$ of 0.025 and a strong $\mat{R}$ threshold of 0.25 with different levels of angular refinement.}
\label{tab:2D_stream_angle_hypre}
\end{table}
\subsubsection{Scattering problem}
\label{sec:scatter}
To test the performance of AIRG with diffusion, we set the total and scattering cross-section to 10.0. Tables \ref{tab:2D_stream_hypre_diffusion} and \ref{tab:2D_stream_hypre_disttwo_diffusion} show that both distance 1 and distance 2 lAIR, respectively, perform similarly, with approximately 3$\times$ growth in WUs from the least to most refined spatial grid. Both the cycle and operator complexities have plateaued though.  

\begin{table}[ht]
\centering
\begin{tabular}{ c c | c c c c c c}
\toprule
CG nodes & NDOFs & $n_\textrm{its}$ & CC & Op Complx &  WUs$^\textrm{full}$ & WUs$^\textrm{DG}$ & Memory\\
\midrule
97 & 2.4\xten{3} & 23 & 2.2 & 1.3 & 77 & 49 & -\\
591 & 1.6\xten{4} & 27 & 3.0 & 1.9 & 112 & 69 & -\\
2313	 & 6.3\xten{4} & 30 & 3.3 & 2.2 & 134 & 82 & - \\
9166	 & 2.5\xten{5} & 34 & 3.4 & 2.2 & 153 & 93 & - \\
35784 & 9.9\xten{5} & 41 & 3.4 & 2.2 & 183 & 110 & - \\
150063 & 4.2\xten{6} & 54 & 3.3 & 2.2 & 238 & 144 & - \\
\bottomrule  
\end{tabular}
\caption{Results from using distance 1 lAIR in \textit{hypre} on a pure scattering problem in 2D with CF splitting by the \textit{hypre} implementation of Falgout-CLJP with a strong threshold of 0.9, drop tolerance on $\mat{A}$ of 1\xten{-4}, $\mat{R}$ of 1\xten{-2} and strong $\mat{R}$ threshold of 0.4.}
\label{tab:2D_stream_hypre_diffusion}
\end{table}

\begin{table}[ht]
\centering
\begin{tabular}{ c c | c c c c c c}
\toprule
CG nodes & NDOFs & $n_\textrm{its}$ & CC & Op Complx &  WUs$^\textrm{full}$ & WUs$^\textrm{DG}$ & Memory\\
\midrule
97 & 2.4\xten{3} & 22 & 2.5 & 1.5 & 80 & 51  & -\\
591 & 1.6\xten{4} & 25 & 3.5 & 2.2 & 117 & 72 & -\\
2313	 & 6.3\xten{4} & 27 & 3.8 & 2.4 & 133 & 81 & - \\
9166	 & 2.5\xten{5} & 31 & 3.7 & 2.4 & 150 & 91 & - \\
35784 & 9.9\xten{5} & 37 & 3.7 & 2.5 & 178 & 108 & - \\
150063 & 4.2\xten{6} & 47 & 3.7 & 2.5 & 225 & 136 & - \\
\bottomrule  
\end{tabular}
\caption{Results from using distance 2 lAIR in \textit{hypre} on a pure scattering problem in 2D with CF splitting by the \textit{hypre} implementation of Falgout-CLJP with a strong threshold of 0.9, drop tolerance on $\mat{A}$ of 1\xten{-4}, $\mat{R}$ of 1\xten{-2} and strong $\mat{R}$ threshold of 0.4.}
\label{tab:2D_stream_hypre_disttwo_diffusion}
\end{table}
AIRG performs simliarly to lAIR in this problem, as shown in Tables \ref{tab:2D_stream_hypre_fc_unconstrained_diffusion} without fixed sparsity and \ref{tab:2D_stream_hypre_fc_diffusion} with fixed sparsity, with around 3$\times$ growth and similar number of WUs. AIRG with fixed sparsity results in slightly lower operator complexities, but both methods result in memory use of around 20 copies of the angular flux; this is higher than that in the streaming limit as the full matrix with scattering and a uniform angular discretisation at one level of refinement has 4$\times$ the nnzs as that in the streaming limit. There is not a great deal of difference in the cycle complexity between AIRG with and without fixed sparsity; this is because the strong $\mat{R}$ threshold of 0.4 results in a very sparse version of $\mat{A}_\textrm{ff}$ being used to construct the GMRES polynomials (and hence very little fill-in relative to the top grid matrix). We can decrease the strong $\mat{R}$ tolerance to decrease the iteration count (in both AIRG and lAIR), but the nnzs in (or equivalently the number of neighbours used to construct) $\mat{Z}$ grows considerably, as might be expected with scattering. 

\begin{table}[ht]
\centering
\begin{tabular}{ c c | c c c c c c c c}
\toprule
CG nodes & NDOFs & $n_\textrm{its}$ & CC & Op. Complx & WUs$^\textrm{full}$ & WUs$^\textrm{DG}$ & Memory\\
\midrule
97 & 2.4\xten{3} & 22 & 1.9 & 1.3 & 68 & 43 & 17.3\\
591 & 1.6\xten{4} & 26 & 2.2 & 2.1 & 88 & 54 & 18.0\\
2313	 & 6.3\xten{4} & 34 & 2.5 & 2.4 & 122 & 74 & 18.7 \\
9166	 & 2.5\xten{5} & 41 & 2.7 & 2.7 & 155 & 94 & 19.5 \\
35784 & 9.9\xten{5} & 45 & 2.8 & 2.8 & 175 & 106 & 19.9 \\
150063 & 4.2\xten{6} & 61 & 2.7 & 2.9 & 232 & 140 & 19.5 \\
\bottomrule  
\end{tabular}
\caption{Results from using AIRG with $m=4$ and without fixed sparsity on a pure scattering problem in 2D with CF splitting by the \textit{hypre} implementation of Falgout-CLJP with a strong threshold of 0.9, drop tolerance on $\mat{A}$ of 1\xten{-4}, $\mat{R}$ of 1\xten{-2} and strong $\mat{R}$ threshold of 0.4.}
\label{tab:2D_stream_hypre_fc_unconstrained_diffusion}
\end{table}

\begin{figure}[th]
\centering
\subfloat[][Solve time]{\label{fig:solve_times_diffusion}\includegraphics[width =0.45\textwidth]{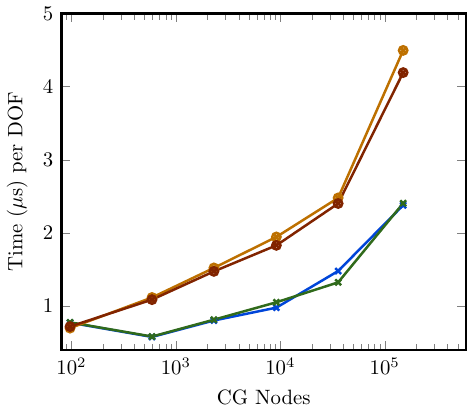}} \quad
\subfloat[][Dashed are time to compute $\hat{\mat{A}}_\textrm{ff}^{-1}$ for AIRG, solid are time to compute $\mat{Z}$]{\label{fig:z_times_diffusion}\includegraphics[width =0.45\textwidth]{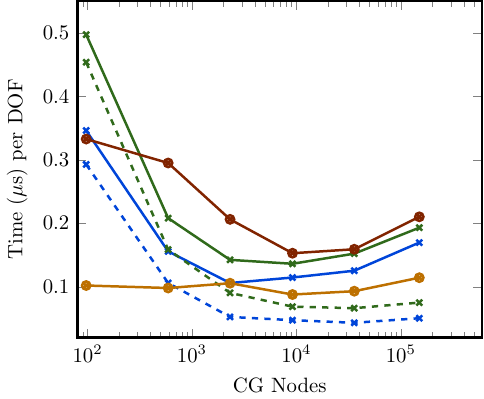}}\\
\subfloat[][Setup time]{\label{fig:setup_times_diffusion}\includegraphics[width =0.45\textwidth]{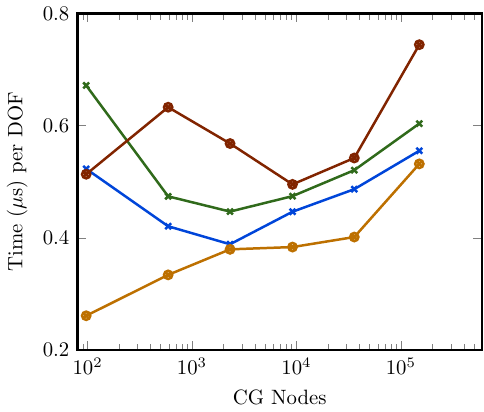}}\quad
\subfloat[][Total time]{\label{fig:total_times_diffusion}\includegraphics[width =0.45\textwidth]{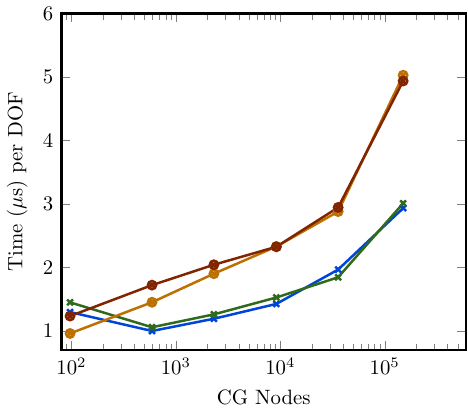}}
\caption{Timings per DOF for AIRG with $m=4$ and lAIR in a 2D pure scattering problem. The \textcolor{matlabblue}{$\times$} is AIRG with fixed sparsity with Falgout-CLJP, the \textcolor{foliagegreen}{$\times$} is AIRG without fixed sparsity and Falgout-CLJP, the \textcolor{deludedorange}{$\otimes$} is distance 1 lAIR with Falgout-CLJP, the \textcolor{fireenginered}{$\otimes$} is distance 2 lAIR with Falgout-CLJP.}
\label{fig:times_diffusion}
\end{figure}

\begin{table}[ht]
\centering
\begin{tabular}{ c c | c c c c c c c c}
\toprule
CG nodes & NDOFs & $n_\textrm{its}$ & CC & Op. Complx & WUs$^\textrm{full}$ & WUs$^\textrm{DG}$ & Memory\\
\midrule
97 & 2.4\xten{3} & 22 & 1.9 & 1.3 & 67 & 43 & 17.3\\
591 & 1.6\xten{4} & 26 & 2.2 & 2.0 & 88 & 54 & 18.0\\
2313	 & 6.3\xten{4} & 34 & 2.5 & 2.4 & 122 & 74 & 18.7 \\
9166	 & 2.5\xten{5} & 38 & 2.7 & 2.7 & 144 & 87 & 19.4 \\
35784 & 9.9\xten{5} & 51 & 2.8 & 2.8 & 196 & 118 & 19.7 \\
150063 & 4.2\xten{6} & 60 & 2.7 & 2.8 & 224 & 135 & 19.3 \\
\bottomrule  
\end{tabular}
\caption{Results from using AIRG with $m=4$ and fixed sparsity on a pure scattering problem in 2D with CF splitting by the \textit{hypre} implementation of Falgout-CLJP with a strong threshold of 0.9, drop tolerance on $\mat{A}$ of 1\xten{-4}, $\mat{R}$ of 1\xten{-2} and strong $\mat{R}$ threshold of 0.4.}
\label{tab:2D_stream_hypre_fc_diffusion}
\end{table}
\fref{fig:times_diffusion} shows the timing results from AIRG and lAIR in this problem. Again we see a curious implementation difference in solve times in \fref{fig:solve_times_diffusion}, as both lAIR and AIRG require simliar number of WUs, but the \textit{hypre} implementation of lAIR requires roughly twice the time to solve. \fref{fig:z_times_diffusion} shows that the time to compute the GMRES polynomial for AIRG is largely constant, with the time to compute $\mat{Z}$ again between distance 1 and distance 2 lAIR; this is also true for the total setup time in \fref{fig:setup_times_diffusion}. \fref{fig:total_times_diffusion} shows that our AIRG implementation is the cheapest method overall, taking around 0.4$\times$ the amount of time as lAIR to solve this problem. If we again equate the solve time between lAIR and AIRG given the similar amount of work required, lAIR and AIRG perform similarly, each requiring approximately 2.9 $\mu$s total time per DOF. 

\begin{figure}[th]
\centering
\subfloat[][Setup time]{\label{fig:setup_times_diffusion_gmres}\includegraphics[width =0.45\textwidth]{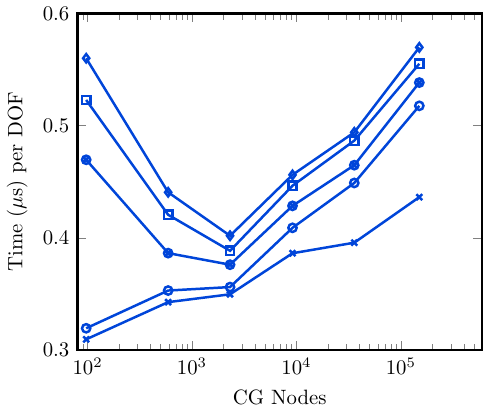}}\quad
\subfloat[][Total time]{\label{fig:total_times_diffusion_gmres}\includegraphics[width =0.45\textwidth]{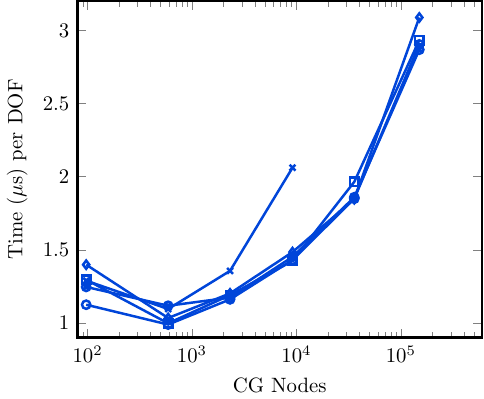}}
\caption{Timings per DOF for AIRG with fixed sparsity, Falgout-CLJP and with varying GMRES polynomial order in a 2D pure scattering problem. The \textcolor{matlabblue}{$\times$} is AIRG with $m=1$, \textcolor{matlabblue}{o} is $m=2$, \textcolor{matlabblue}{$\otimes$} is $m=3$, \textcolor{matlabblue}{$\square$} is $m=4$, \textcolor{matlabblue}{$\diamond$} is $m=5$.}
\label{fig:times_diffusion_gmres}
\end{figure}

\fref{fig:times_diffusion_gmres} shows the results from changing the GMRES polynomial order and similar trends to that in the streaming limit can be seen, namely the 0th order polynomial is very cheap to setup, but results in the highest total time. Indeed the 0th order polynomial did not converge at the two highest spatial refinements. The first through fourth order polynomials all result in similar total times, with 62, 61, 60 and 64 iterations, respectively. This result indicates that the higher polynomial order does not necessarily help decrease the iteration count with scattering. This is because the strong $\mat{R}$ threshold is so high; decreasing this makes the effect of the polynomial order (and the fixed sparsity) much more pronounced, but we found the lowest overall total times by allowing heavy dropping. Given using the full matrix is not scalable with angular refinement, we don't show convergence results in that case; instead the next section uses the iterative method defined in \secref{sec:iterative} on this scattering problem. 

\subsection{Additively preconditioned iterative method}
\label{sec:mf_iterative}
In this section we show the performance of the iterative method from \secref{sec:iterative} in the scattering limit. The results in this section are not designed to show the performance of a standard DSA method with different scattering ratios, optical cell lengths, etc; we appeal to the wealth of literature on the topic. Instead we wish to show that the additive combination of our preconditioners is effective and that multigrid methods can be used to invert these operators scalably. As discussed, this means forming the streaming/removal operator $\mat{M}_\Omega$, a CG diffusion operator $\mat{D}_\textrm{diff}$ and the streaming/removal components $\mat{B}_\Omega$ and/or $\mat{C}_\Omega$, all of which can be done scalably. We use 1 V-cycle of AIRG with the same drop/strong tolerances as in \secref{sec:pure_stream} to apply $\mat{M}_\Omega^{-1}$ and 1 V-cycle of \textit{boomerAMG} (with default options) to apply $\mat{D}_\textrm{diff}^{-1}$ per outer GMRES iteration. 
\begin{figure}[ht]
\centering
\subfloat[][Streaming operator]{\label{fig:streaming}\includegraphics[width =0.45\textwidth]{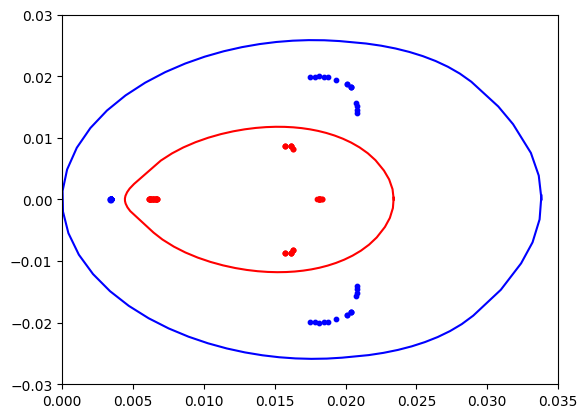}}\quad
\subfloat[][Streaming/removal operator with a total cross-section of 10.0.]{\label{fig:streaming_removal}\includegraphics[width =0.45\textwidth]{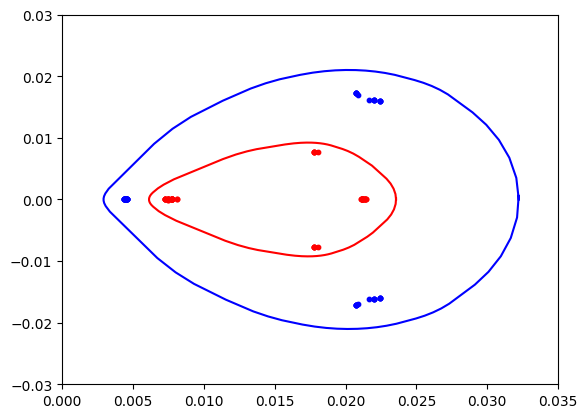}}
\caption{The 10 biggest and smallest eigenvalues (by real, imaginary parts and magnitude) (dots) and field of values (solid lines) of different operators, with the \textcolor{red}{red} the equivalent for $\mat{A}_\textrm{ff}$ with CF splitting by Falgout-CLJP. Computed on the third refined spatial grid with level one angular refinement.}
\label{fig:spectrum_stream_removal}
\end{figure}

For our iterative method to be effective, 1 V-cycle of both methods must reduce the error by a fixed amount with space/angle refinement (which is equivalent to a solve with a fixed tolerance taking a fixed amount of work). We can assume that multigrid methods such as \textit{boomerAMG} can invert the diffusion operator with fixed work (we also scale the diffusion operator by its inverse diagonal prior to use), but in \secref{sec:pure_stream} we only showed that AIRG can invert the streaming operator with fixed work in the solve, rather than the streaming/removal operator. Thankfully the removal term results in a better conditioned matrix given the extra term on the (block) diagonals; the streaming limit is the most difficult to solve. \fref{fig:spectrum_stream_removal} shows (part of) the spectrum of the streaming operator vs the streaming/removal operator for the 2D source problem with the third refined grid, level one angular refinement and total cross-section of 10.0. We can see that the smallest eigenvalues of the streaming/removal operator are (slightly) further from the origin.

The convergence of AIRG relies on the convergence of our GMRES polynomial approximations to $\mat{A}_\textrm{ff}^{-1}$. We can also see in \fref{fig:spectrum_stream_removal} that the eigenvalues of $\mat{A}_\textrm{ff}$ are more compact than that of the full operators, confirming that the CF splitting is helping produce a better conditioned $\mat{A}_\textrm{ff}$ in both cases. We know that our operators are non-normal, so the spectrum does not completely determine the convergence of our GMRES polynomials \cite{Meurant2020}. Given this, \fref{fig:spectrum_stream_removal} also plots the field of values (a.k.a., the numerical range) of our operators, given by
\begin{equation}
\mathcal{F}(\mat{A}) = \{ \mat{x}^* \mat{A} \mat{x} \, | \, \mat{x}\mat{x}^*=1, \mat{x}\in \mathbb{C}^n\},
\label{eq:fov}
\end{equation}
which is a convex set that contains the eigenvalues. To give some insight into the convergence of the GMRES polynomials, we define $\mu$ to be the distance from the origin, or
\begin{equation}
\mu = \min_{z \in \mathcal{F}(\mat{A})} |z|.
\end{equation} 
\cite{Beckermann2005} show that (see also \cite{Meurant2020}) if the field of values doesn't contain the origin, $\beta \in (0, \pi/2)$ such that $\cos(\beta)=\mu/||\mat{A}||$ and the Hermitian part of $\mat{A}$, namely $(\mat{A}+\mat{A}^*)/2$ is positive definite then the residual at step $m$ is bounded by 
\begin{equation}
||\mat{r}^m|| \leq ||\mat{r}^0|| \left(2 + \frac{2}{\sqrt{3}}\right) (2 + \gamma_\beta) \gamma_\beta^m, 
\label{eq:gmres_bound}
\end{equation}
where 
\begin{equation}
\gamma_\beta = 2 \sin \left( \frac{\beta}{4 - 2 \beta /\pi} \right).
\label{eq:gamma_beta}
\end{equation}
We confirmed numerically that none of our operators or their fine-fine sub-matrices touch the origin (and hence the field of values are all in the right-half of the complex plane) and that their Hermitian parts are positive definite. 

For the $\mat{A}_\textrm{ff}$ component of the streaming operator on the top grid, pictured in \fref{fig:spectrum_stream_removal} we found that with $m=4$, \eref{eq:gmres_bound} gives $||\mat{r}^m|| \leq 9.41 ||\mat{r}^0||$, while for the $\mat{A}_\textrm{ff}$ component of the streaming/removal operator we have $||\mat{r}^m|| \leq 9.40 ||\mat{r}^0||$ (the disk bound in \cite{Liesen2020} gives a similar conclusion). These bounds are not particularly tight, but they do indicate that a 3rd order GMRES polynomial should result in a smaller residual for the streaming/removal operator and hence we would expect AIRG to perform better.

We should note that as $\mu \rightarrow 0$, $\beta$ gets closer to $\pi/2$ and the asymptotic convergence factor, $\gamma_\beta^m \rightarrow 1$. In general this means the further the minimum field of values is from the origin, the better the convergence; this is also demonstrated by considering the disk bound in \cite{Liesen2020}, given by $|\delta/c|=(1-\cos(\beta))/(1+\cos(\beta)) < \gamma_\beta$, where $\delta$ and c are the radius and centre of a disk, respectively, that covers $\mathcal{F}(\mat{A})$. This helps explain why using GMRES polynomials to approximate $\mat{A}_\textrm{ff}$ can be effective even when GMRES polynomial preconditioning of the full operators may not be; \fref{fig:streaming} shows that the field of values for the streaming operator almost touches the origin, with $\mu \approx 3.2\xtenm{-5}$ and hence single-level GMRES polynomial preconditioning would likely not be effective in this problem (this is backed by numerical experiments; we find considerable growth in the iteration count with refinement). This hints at the importance of combining GMRES polynomials with a reduction multigrid. 

\begin{table}[ht]
\centering
\begin{tabular}{ c c | c c c c c c c c}
\toprule
CG nodes & NDOFs & $n_\textrm{its}$ & CC & Op. Complx & WUs$^\textrm{mf}$ & WUs$^\textrm{DG}$ & Memory\\
\midrule
97 & 2.4\xten{3} & 23 & 3.0 & 1.6 & 33 & 61 & 16.5\\
591 & 1.6\xten{4} & 24 & 4.0 & 1.0 & 36 & 67 & 17.2\\
2313	 & 6.3\xten{4} & 25 & 4.1 & 1.7 & 38 & 69 & 17.2 \\
9166	 & 2.5\xten{5} & 26 & 4.2 & 2.4 & 39 & 72 & 17.2 \\
35784 & 9.9\xten{5} & 26 & 4.4 & 2.8 & 39 & 73 & 17.3 \\
150063 & 4.2\xten{6} & 26 & 4.6 & 3.2 & 39 & 73 & 17.5 \\
\bottomrule  
\end{tabular}
\caption{Results from using additive preconditioning on a pure scattering problem with total and scattering cross-section of 10.0 in 2D. The cycle and operator complexity listed are for AIRG on $\mat{M}_\Omega$ with CF splitting by Falgout-CLJP.}
\label{tab:additive}
\end{table}
\begin{table}[ht]
\centering
\begin{tabular}{ c c c | c c c c c c c c}
\toprule
CG nodes & Angle lvl. & NDOFs & $n_\textrm{its}$ & CC & Op. Complx & WUs$^\textrm{mf}$ & WUs$^\textrm{DG}$ & Memory\\
\midrule
2313 & 1 & 6.3\xten{4} & 25 & 4.1 & 1.7 & 38 & 69 & 17.2 \\
2313 & 2 & 2.5\xten{5} & 27 & 4.0 & 1.4 & 40 & 71 & 16.5 \\
2313 & 3 & 1\xten{6} & 28 & 4.1 & 1.4 & 41 & 73 & 16.2 \\
\bottomrule  
\end{tabular}
\caption{Results from using additive preconditioning on a pure scattering problem with total and scattering cross-section of 10.0 in 2D with angle refinement. The cycle and operator complexity listed are for AIRG on $\mat{M}_\Omega$ with CF splitting by Falgout-CLJP.}
\label{tab:additive_angle}
\end{table}
\begin{table}[ht]
\centering
\begin{tabular}{ c c | c c c c c c c c}
\toprule
CG nodes & NDOFs & $n_\textrm{its}$ & CC & Op. Complx & WUs$^\textrm{mf}$ & WUs$^\textrm{DG}$ & Memory\\
\midrule
97 & 2.4\xten{3} & 18 & 3.6 & 1.9 & 27 & 50 & 17.1\\
591 & 1.6\xten{4} & 18 & 4.0 & 2.4 & 27 & 50 & 17.2\\
2313	 & 6.3\xten{4} & 18 & 4.3 & 2.8 & 28 & 51 & 17.4 \\
9166	 & 2.5\xten{5} & 19 & 4.5 & 3.1 & 29 & 54 & 17.5 \\
35784 & 9.9\xten{5} & 19 & 4.7 & 3.3 & 30 & 55 & 17.6 \\
150063 & 4.2\xten{6} & 19 & 4.8 & 3.5 & 30 & 55 & 17.7 \\
\bottomrule  
\end{tabular}
\caption{Results from using additive preconditioning on a pure scattering problem with total and scattering cross-section of 1.0 in 2D. The cycle and operator complexity listed are for AIRG on $\mat{M}_\Omega$ with CF splitting by Falgout-CLJP.}
\label{tab:additive_one}
\end{table}
\begin{table}[ht]
\centering
\begin{tabular}{ c c c | c c c c c c c c}
\toprule
CG nodes & Angle lvl. & NDOFs & $n_\textrm{its}$ & CC & Op. Complx & WUs$^\textrm{mf}$ & WUs$^\textrm{DG}$ & Memory\\
\midrule
2313 & 1 & 6.3\xten{4} & 18 & 4.3 & 2.8 & 28 & 51 & 17.4 \\
2313 & 2 & 2.5\xten{5} & 18 & 4.3 & 2.8 & 27 & 49 & 16.7 \\
2313 & 3 & 1\xten{6} & 19 & 4.3 & 2.8 & 29 & 51 & 16.5 \\
\bottomrule  
\end{tabular}
\caption{Results from using additive preconditioning on a pure scattering problem with total and scattering cross-section of 1.0 in 2D with angle refinement. The cycle and operator complexity listed are for AIRG on $\mat{M}_\Omega$ with CF splitting by Falgout-CLJP.}
\label{tab:additive_one_angle}
\end{table}

We see in \tref{tab:additive} our additively preconditioned iterative method is effective with a total and scatter cross-section of 10, with the iteration count growing from 23 to a plateau of 26 with spatial refinement. The work is very close to constant, with fixed iteration count and slight growth in the cycle complexity, even though we used AIRG with fixed sparsity. As such we didn't investigate using AIRG without fixed sparsity, as the fixed sparsity was sufficient to give plateauing work. This helps confirm the observations above, namely that AIRG is more effective on the streaming/removal operator. 

Comparing to the results in \secref{sec:scatter}, it uses roughly 73 DG WUs, compared to around 135 when using either AIRG or lAIR as a preconditioner on the full matrix. It also uses less memory at approximately 18 copies of the angular flux, even though we have to store the diffusion operator and several extra temporary vectors. Compared to the pure streaming problem, this method requires approximately 4.1$\times$ more work; we can see in \tref{tab:additive} that this work largely comes from computing the matrix-free matvec with scattering, with 26 iterations at the highest spatial refinement requiring 39 WUs$^\textrm{mf}$. The split of work is 26 WUs$^\textrm{mf}$ in the matvec required by the outer GMRES, 11 WUs$^\textrm{mf}$ to apply the additive preconditioners and around 2 WUs$^\textrm{mf}$ to compute the source and $\bm{\Theta}$. Similarly, \tref{tab:additive_one} shows a (lower) constant iteration count with a lower total and scatter cross-section of 1.0. 

Given the streaming/removal operator is easier to solve, lAIR performs better when used additively to invert $\mat{M}_\Omega$, compared with just the streaming operator. With a total and scatter cross-section of 10.0, we find 1 V-cycle of both distance 1 and 2 lAIR give 30, 33, 31, 34, 36 and 39 iterations with spatial refinement, with similar grid and operator complexities to AIRG. Increasing the number of FCF smooths from 1 to 3 results in an iteration count with less growth, namely 30, 25, 25, 26, 26 and 27 iterations, but the cycle complexity at the finest level of refinement is large at 10.7, compared with AIRG at 4.6. We also know from \secref{sec:pure_stream} that the iteration count of lAIR grows in the streaming limit, so we do not test lAIR any further as part of our additive method. 

Tables \ref{tab:additive_angle} and \ref{tab:additive_one_angle} show that the additive method with AIRG and angular refinement perform well, as the iteration count and the work required is very close to constant. Importantly we can also see the memory use is fixed. These results show that as might be expected, using an (inconsistent) CG DSA can form an effective preconditioner in scattering problems when used with an outer GMRES iteration. Importantly the combination of a single V-cycle of AIRG used on the streaming/removal operator and a single V-cycle of a traditional multigrid on the diffusion operator can be used additively and results in almost constant work with spatial and angular refinement on unstructured grids.
\section{Conclusions}
This paper presented a new reduction multigrid based on approximate ideal restrictors (AIR) combined with GMRES polynomials (AIRG) with excellent performance in advection-type problems. Matrix polynomial methods have been used for many years in multilevel methods but we believe we are the first to use GMRES polynomials in this fashion. Reduction multigrids and LDU methods in particular benefit from using GMRES polynomials, as the improved conditioning of $\mat{A}_\textrm{ff}$, when compared to $\mat{A}$, can allow the formation of good approximate inverses with low polynomial orders. This allowed us to easily build both approximate ideal restrictors, approximate ideal prolongators (without the need to compute near-nullspace vectors) and perform F-point smoothing (without the need to compute additional dampening parameters).

GMRES polynomials share many advantages with other polynomial methods; in particular their coefficients can be computed very simply; low-order polynomials don't require additional work to ensure stability (like in \cite{Liu2015, Loe2021}); explicitly forming approximate matrix inverses is simple and only involves matrix-matrix products or if desired; the polynomials can be applied matrix-free; their application is highly parallel with their setup able to use communication-avoiding techniques; and they also work well across a range of symmetric and asymmetric problems. 

When applied to the time independent Boltzmann Transport Equation (BTE) we could solve pure streaming problems (i.e., in the pure advection limit) on unstructured spatial grids with space/angle refinement with fixed memory use. The time-independent streaming limit is the most challenging to solve and we found we could either get fixed work in the solve and growth in the setup, or by introducing fixed sparsity into the matrix-powers of our GMRES polynomials, we found fixed work in the setup with growth in the solve. We found good performance from using between first to fourth order GMRES polynomials on each level of our multigrid. Fixing the sparsity of our third-order ($m=4$) GMRES polynomials resulted in a fixed FLOP count in the setup, and building an implementation of a matmatmult $\mat{A}=\mat{B}\mat{C}$ where the three matrices share the same sparsity would reduce the implementation costs of our setup for second order polynomials and higher. With fixed sparsity we found at most 20\% growth in the work to solve with either 6 levels of spatial refinement or three levels of uniform angular refinement.

A balance must be struck between the scalability of the solve vs expense of the setup, but we believe this is the first method to show scalable solves with a stable spatial discretisation that doesn't feature lower-triangular structure in the streaming limit of the BTE. We did not spend much effort tweaking parameters and we have found that that we can get better performance in these problems with standard AMG tweaks such as level specific drop tolerances. 

We also compared AIRG to two different reduction multigrids and found performance advantages; one where sparse approximation inverses (SAIs) are used to approximate $\mat{A}_\textrm{ff}^{-1}$ instead of our GMRES polynomials, and the lAIR implementation in \textit{hypre}. We used \textit{ParaSails} in \textit{hypre} to form SAIs with the fixed sparsity of $\mat{A}_\textrm{ff}$ and found almost identical convergence behavior to fixed sparsity AIRG (and hence the same solve time) in the streaming limit. We found however that the setup of the SAIs took twice as long as our GMRES polynomials with $m=4$. For lAIR, we could not find a set of parameters that resulted in fixed work in the solve. Our further investigations suggest the combination of distance 3 or 4 lAIR plus (only) F-point smooths are required to get scalable results with lAIR, but this is not practical given the setup/communication costs. 

In comparison to distance 1 or 2 lAIR, AIRG took roughly two to three times less work to solve. Timing the setup showed that computing $\mat{Z}$ with our third-order GMRES polynomial approximations cost between that of computing $\mat{Z}$ with distance 1 and 2 lAIR. The total time of our setup at the highest level of spatial refinement matched that of distance 2 lAIR. The total time (setup plus solve) for AIRG was roughly $3\times$ less than lAIR, though implementation differences make this comparison difficult. We then investigated using AIRG and lAIR on the full matrix formed with scattering. Forming this matrix cannot be done scalably with angular refinement, but we showed that AIRG is applicable in the diffuse limit, performing about as well as lAIR. 

We then built an iterative method that used the additive combination of two preconditioners applied to the angular flux; 1 V-cycle of AIRG was used to invert the streaming/removal operator and 1 V-cycle of \textit{boomerAMG} was used to invert a CG diffusion operator (i.e., an inconsistent DSA). The streaming/removal operator is easier to solve than the streaming operator and hence we found the work in the solve plateaued with fixed sparsity AIRG. Using distance 1 or 2 lAIR to invert the streaming/removal operator resulted in cycle complexities over twice that of fixed sparsity AIRG. Given the performance shown here it would be worth investigating the use of AIRG as part of a standard DG FEM source iteration; preliminary work reveals AIRG performs similarly when used with DG streaming or streaming/removal operators.

The only remaining consideration is how we can apply this method with our previously developed angular adaptivity and the parallel performance, which we will investigate in future work. AIRG should be performant in parallel, as the entire multigrid hierarchy can be applied with only matrix-vector products (i.e., no reductions). The CF splitting algorithm used has a parallel implementation available in \textit{hypre}. The GMRES polynomial coefficients on each level must be computed once during the setup and can be trivially stored for multiple solves. Furthermore given our use of low-order GMRES polynomials in AIRG, we found a single step method based on a QR factorisation of the power basis could be used to generate these coefficients stably. In parallel we could therefore use a tall-skinny QR and generate the coefficients of a polynomial of order $m-1$ with $m$ matvecs and a single all-reduce on each level. For zero and first order polynomials, there is no other communication required. For second order and higher, the remainder of the GMRES polynomial setup uses $m-2$ matmatmults and matmatadds to compute matrix-powers. If we impose the aforementioned fixed sparsity we only need to communicate the required off-processor rows of $\mat{A}_\textrm{ff}$ in the matmatmults once in order to compute those matrix powers, regardless of the order of the polynomial. 

Given the results in this paper, we believe the combination of our low-memory sub-grid scale discretisation, AIRG with low-order GMRES polynomials, and an iterative method that additively preconditions with the streaming/removal operator and an inconsistent CG DSA forms an excellent method for solving transport problems on unstructured grids in both the streaming and scattering limit.
\section*{Acknowledgments}
The authors would like to acknowledge the support of the EPSRC through the funding of the EPSRC grants EP/R029423/1 and EP/T000414/1.




\section*{References}
\bibliographystyle{model1-num-names}
\bibliography{bib_library}







\end{document}